\documentclass[reprint,superscriptaddress,amsmath,amssymb,aps,pra]{revtex4-1}

\usepackage{graphicx}
\usepackage{dcolumn}
\usepackage{bm}
\usepackage[dvipsnames]{xcolor}

\usepackage{tabularx}
\newcolumntype{Y}{>{\centering\arraybackslash}X}

\newcommand{\threejm}[6]{ \left(\begin{array}{ccc} #1 & #3 & #5\\
                                              #2 & #4 & #6
                                \end{array}
                          \right)}
\newcommand{\sixj}[6]{ \left\{\begin{array}{ccc} #1 & #2 & #3\\
                                              #4 & #5 & #6
                                \end{array}
                          \right\}}
\newcommand{\ninej}[9]{ \left\{\begin{array}{ccc} #1 & #2 & #3\\
                                              #4 & #5 & #6\\
                                              #7 & #8 & #9\\
                                \end{array}
                          \right\}}

\begin{document}

\title{\emph{Ab initio} calculation of the spectrum of Feshbach resonances in NaLi + Na collisions}

\author{Tijs Karman}
\affiliation{Institute for Molecules and Materials, Radboud University, Heijendaalseweg 135, 6525 AJ Nijmegen, the Netherlands}

\author{Marcin Gronowski}
\affiliation{Faculty of Physics, University of Warsaw, Pasteura 5, 02-093 Warsaw, Poland}

\author{Micha{\l} Tomza}
\affiliation{Faculty of Physics, University of Warsaw, Pasteura 5, 02-093 Warsaw, Poland}

\author{Juliana J. Park}
\affiliation{Research Laboratory of Electronics, MIT-Harvard Center for Ultracold Atoms,
Department of Physics, Massachusetts Institute of Technology, Cambridge, Massachusetts 02139, USA}

\author{Hyungmok Son}
\affiliation{Research Laboratory of Electronics, MIT-Harvard Center for Ultracold Atoms,
Department of Physics, Massachusetts Institute of Technology, Cambridge, Massachusetts 02139, USA}
\affiliation{Department of Physics, Harvard University, Cambridge, Massachusetts 02138, USA}

\author{Yu-Kun Lu}
\affiliation{Research Laboratory of Electronics, MIT-Harvard Center for Ultracold Atoms,
Department of Physics, Massachusetts Institute of Technology, Cambridge, Massachusetts 02139, USA}

\author{Alan O. Jamison}
\affiliation{Institute for Quantum Computing and Department of Physics \& Astronomy,
University of Waterloo, Waterloo, Ontario N2L 3G1, Canada}

\author{Wolfgang Ketterle}
\affiliation{Research Laboratory of Electronics, MIT-Harvard Center for Ultracold Atoms,
Department of Physics, Massachusetts Institute of Technology, Cambridge, Massachusetts 02139, USA}

\date{\today}

\begin{abstract}
We present a combined experimental and theoretical study of the spectrum of magnetically tunable Feshbach resonances in NaLi $(a^3\Sigma^+)$ $+$ Na collisions.
In the accompanying paper, we observe experimentally 8 and 17 resonances occur between $B=0$ and $1400$~G in upper and lower spin-stretched states, respectively.
Here, we perform \emph{ab initio} calculations of the NaLi $+$ Na interaction potential and describe in detail the coupled-channel scattering calculations of the Feshbach resonance spectrum.
The positions of the resonances cannot be predicted with realistic uncertainty in the state-of-the-art \emph{ab initio} potential,
but our calculations yield a \emph{typical} number of resonances that is in near-quantitative agreement with experiment. We  show that the main coupling mechanism results from spin-rotation and spin-spin couplings in combination with the anisotropic atom-molecule interaction. The calculations furthermore explain the qualitative difference between the numbers of resonances in either spin state.
\end{abstract}

\maketitle

\section{Introduction}

Magnetically tunable resonances have become an indispensable tool for controlling the interactions between ultracold atoms \cite{chin:10}.
Control over the interactions enables using atoms as quantum simulators to study for example condensed matter physics \cite{bloch:12,bakr:10,struck:11,simon:11,chiu:18}.
Feshbach resonances are also used to associate pairs of ultracold atoms into weakly bound molecules \cite{kohler:06}.
These weakly bound molecules can be transferred coherently to their absolute ground state using a stimulated Raman adiabatic passage (STIRAP) \cite{bergmann:98},
which has become a common tool for producing ultracold ground-state molecules \cite{ni:08,takekoshi:14,molony:14,guo:16}.
Resonances in molecular collisions could similarly enable the control of contact interactions between molecules, or between molecules and atoms.
This could enable sympathetic cooling of molecules \cite{son:20}, and the bottom up construction of ultracold polyatomic molecules \cite{yang:22}.
Furthermore, scattering resonances are incredibly sensitive to the details of the interaction potential and provide an interesting testing ground for theory \cite{gronowski:20}.
In particular, there is an active debate about the nature of so-called ``sticky collisions'' between ultracold molecules \cite{mayle:12,mayle:13,christianen:19b},
their role in collisional losses \cite{christianen:19a},
and what simplified statistical models can be used to describe these \cite{mayle:12,mayle:13,croft:23,christianen:21}.
Observation of atom-molecule and molecule-molecule resonances could directly test some of these models by measuring the ``effective'' density of states \cite{christianen:19b,man:22},
the probability of short-range loss \cite{croft:20,christianen:21},
and by probing the mechanism of this loss.

Collisional loss also represents a major hurdle in the observation of Feshbach resonances.
The resonance states correspond to quasi-bound states of the atom-molecule or molecule-molecule collision complex.
In the presence of collisional loss, however, these states experience decay and the resonances broaden \cite{christianen:21}.
In the case of so-called universal loss \cite{idziaszek:10}, which is consistent with many experimental observations of molecule-molecule \cite{ni:08,danzl:10,takekoshi:14,molony:14,park:15,guo:16,rvachov:17,seesselberg:18,yang:19} loss rates, the short-range loss becomes complete and the resonances disappear entirely.
Universal loss is also observed for some but not all ultracold molecule-atom collisions \cite{gregory:21,wang:21c,nichols:22,voges:22}.
For these reasons it is an open question for which systems scattering resonances exist.  

Recently, scattering resonances have been observed in NaK+K \cite{yang:19},
NaLi+Na \cite{son:22},
and even NaLi+NaLi \cite{park:23} collisions, but the three systems are very different.
The resonances observed in NaK+K could be interpreted as long-range states that are further split into several resonances by hyperfine interactions,
and which could tentatively be assigned free-molecule and free-atom quantum numbers \cite{wang:21c}.
In the case of collisions between triplet NaLi molecules and Na atoms in the spin stretched state, the background loss is much less than universal.
This situation, however, could be unique to triplet NaLi, which is the lightest bialkali realized that furthermore has relatively weak interactions with Na in the spin-stretched state.
By contrast, NaLi bimolecular collisions exhibit universal background loss.
Nevertheless, a single resonance has been observed in NaLi+NaLi collisions,
though assignment of the observed resonance in terms of a stable resonance state is complicated \cite{park:23}.

Here, we present a combined experimental and theoretical study of the spectrum of magnetically tunable Feshbach resonances in NaLi $(a^3\Sigma^+)$ + Na collisions.
In the accompanying paper we show 8 and 17 resonances occur between $B=0$ and $1400$~G in the upper and lower spin-stretched states, respectively \cite{companion}.
We perform \emph{ab initio} calculations of the NaLi+Na interaction potential,
and use these to perform coupled-channel scattering calculations of the Feshbach resonance spectrum.
We show that the positions of the resonances cannot be predicted with realistic uncertainty in the state-of-the-art \emph{ab initio} potential,
but that our calculations can nevertheless confirm the expected number of resonances in this magnetic field range.
Furthermore, we explain the qualitative difference between the numbers of resonances in either spin state,
and the nature of the coupling mechanism.
In this paper, we describe in detail how these calculations were performed,
and also present additional supporting calculations of the density of states and the effect of hyperfine interactions on the spectrum of resonances.

\section{Interaction potential \label{sec:interaction}}

Intermolecular interactions in the triatomic NaLi+Na system can be formally decomposed into a pairwise additive two-body part and a pairwise nonadditive three-body part.
The two-body part is the sum of interatomic interaction between all atomic pairs obtained neglecting the presence of other atoms. The nonadditive three-body interaction results from the electron density modification and correlation extending beyond atomic pairs and can also be understood as the change of pairwise interactions due to the presence of other atoms.

\subsection{Two-body interactions}

We first consider the two-body interactions.
The interaction between two alkali-metal atoms $A$ and $B$ depends on the distance between the atoms and whether the spins are singlet or triplet coupled,
which can be represented as
\begin{align}
\hat{V}_{AB}(r_{AB}) = \bar{V}_{AB}(r_{AB}) + \hat{s}_A\cdot\hat{s}_B \Delta V_{AB}(r_{AB}),
\label{eq:twobody}
\end{align}
where $\bar{V}_{AB}(r) = [3V_{AB}^{(T)}(r)+V_{AB}^{(S)}(r)]/4$ and $\Delta V_{AB}(r)=V_{AB}^{(T)}(r) - V_{AB}^{(S)}(r)$.
The singlet $V_{AB}^{(S)}$ and triplet $V_{AB}^{(T)}$ potentials for NaNa and NaLi are taken from Refs.~\cite{knoop:11,steinke:12,rvachov:18}.
These are empirical potentials that reproduce the known molecular spectroscopy and atom-atom scattering lengths accurately.
The total additive electronic interaction is then computed for fixed molecular bond length,
as a function of the molecule-atom center-of-mass separation $R$ and the Jacobi angle $\theta$ as
\begin{equation}
\hat{V}^{2b}(R,\theta)=\sum_{A<B}\hat{V}_{AB}(r_{AB}),
\end{equation}
where, the three interatomic distances $r_{AB}$ are obtained as a function of $R$ and $\theta$ and evaluated with Eq.~\eqref{eq:twobody} for each pair. 
This total additive interaction is represented in a Legendre expansion convenient for scattering calculations
\begin{align}
\hat{V}^{2b}(R,\theta) &= \sum_L P_L(\cos\theta) \times \Big[ V^{(0)}_L(R) + \hat{s}_1\cdot\hat{s}_2 V^{(12)}_L(R)\nonumber \\
&+ \hat{s}_1\cdot\hat{s}_3 V^{(13)}_L(R) + \hat{s}_2\cdot\hat{s}_3 V^{(23)}_L(R) \Big],
\label{eq:expansion}
\end{align}
computed by Gauss-Legendre quadrature including terms up to $L=70$.
The required matrix elements in the channel basis are given in the Appendix.

\subsection{Nonadditive three-body potential}
The nonadditive three-body contribution to the interaction energy in the triatomic $ABC$ system is computed as
\begin{equation}\label{eq:ABC}
V^{3b}(R,\theta)=E_{ABC}-\sum_{A<B}E_{AB}+\sum_A E_A\,,
\end{equation}
where the total energies of the trimer $E_{ABC}$, dimers $E_{AB}$, and monomers $E_{A}$ at the trimer geometry $(R,\theta)$ are computed using a trimer basis set. 

First, the total energies and resulting three-body term are computed using the coupled cluster method~\cite{Musial2007} with the full treatment of single and double excitations and estimation of the connected triples contribution non-iteratively by many-body perturbation theory, CCSD(T),~\cite{KHW93}. Next, the correction to account for the remaining triple excitations in the coupled-cluster expansion, CCSDT~\cite{Kallay2001}, is added. Large Gaussian basis sets are employed. Thus, all total energies in Eq.~\eqref{eq:ABC} are calculated using 
\begin{equation}
E=E^\text{HF}_\text{apCV5Z}+\delta E^\text{CCSD(T)}_\text{CBS(Q,5)}+\delta E^\text{CCSDT}_\text{apCVTZ}\,,
\end{equation}
where the Hartree-Fock energy, $E^\text{HF}_\text{apCV5Z}$, is calculated in the Douglas-Kroll correlation-consistent polarized core-valence quintuple-$\zeta$ quality basis sets, aug-cc-pCV5Z-DK~\cite{bas1}. Next, the correlation energy at the CCSD(T) level, $\delta E^\text{CCSD(T)}_\text{CBS(Q,5)}$, is extrapolated to the complete basis set (CBS) limit with the two-point formula~\cite{L3} and the aug-cc-pCVQZ-DK and aug-cc-pCV5Z-DK basis sets. Relativistic effects are included in those calculations with eXact-2-Component (X2C) Hamiltonian~\cite{X2C2012}. Finally, the full triples correction $\delta E^\text{CCSDT}_\text{apCVTZ}$ defined as a difference between CCSDT and CCSD(T) results is calculated using the aug-cc-pCVTZ basis sets. We correlate only the three valence electrons in the CCSDT calculations, which in the case of the NaLi+Na system is equivalent to reaching the full configuration interaction (FCI) quality for valence electrons.

The CCSD(T) calculations are performed with the MOLPRO package of ab initio programs~\cite{MOLPRO-WIREs,MOLPRO}, while the CCSDT results are obtained with the MRCC2019 software~\cite{MRCCjcp,MRCC}.
The Legendre expansion of the nonadditive three-body interaction potential is available in the Supplementary Material.

The nonadditive three-body interaction potential is computed for the spin-stretched quartet state where the coupled-cluster method can be applied.
We subsequently assume the nonadditive part of the interaction is spin-independent and use it for the doublet states, too. This approximation may be justified by much higher importance of the three-body interaction for the spin-stretched state where it is larger than the two-body contribution, while deeply bound doublet states are strongly dominated by two-body interactions. Additionally, the decomposition into additive and nonadditive parts is ambiguous for the relevant low-spin states.
Finally, later we will show that the details of doublet states are less important for resonance prediction.
We note that the spin-dependence of the two-body interactions is fully accounted for.

\subsection{Electron spin-spin interaction}

The electron spin-spin interaction can originate either from direct spin–spin magnetic dipole–dipole interaction or indirect interaction in the second-order of perturbation theory mediated by spin-orbit coupling. The total effective Hamiltonian for the electron spin-spin interaction reads as 
\begin{equation}
\hat{H}_{SS}=\bm{\hat{S}}^T \bm{D} \bm{\hat{S}}
\end{equation}
where $\bm{\hat{S}}$ represents spin of the system and $\bm{D}$ is $3\times3$ matrix. In the magnetic axis frame the $\bm{D}$ become diagonal, and the effective Hamiltonian can be parameterized by $D$ and $E$ constants \cite{Jaszunski2017}, as
\begin{equation}
\hat{H}_{SS}=D\left[ \hat{S}_{z}^2 - \frac{1}{3}S(S+1) \right]\ + E \left[ \hat{S}_x^2 - \hat{S}_y^2 \right]
\end{equation}
where $\hat{S}_x$ is the $x$ component of the spin operator $\hat{\bm{S}}$ with respect to the magnetic axis system, and similar for $y$ and $z$. The magnetic axes depend on the geometry of the NaLi+Na complex. In the case of linear configurations, the $z$ axis coincides with the NaLi axis. In the global minima of the potential energy surface, the $z$ is perpendicular to the plane defined by atoms, whereas the angle between $y$ and the NaLi axis is about 30 degrees.
Preliminary computations with multi-reference configuration interaction methods show that the spin-orbit interaction plays a minor role here, and $E$ is orders of magnitude smaller than $D$. Thus, we neglect the spin-orbit interaction, include only two-electron spin–spin direct magnetic interaction. We use the multi-reference averaged quadratic coupled-cluster (MR-AQCC)~\cite{SZALAY1993} electronic wave function as implemented~\cite{GanyushinNeese2006} in Orca~\cite{orca1,orca2}. We describe scalar relativistic effects by the Douglass-Kroll-Hess (DKH) Hamiltonian~\cite{Dyall2000} and include the picture-change effects~\cite{SandhoeferNeese2012}. The aug-cc-pCVTZ-DK basis sets are used. 
The components of the $\bm{D}$ matrix, the coupling coefficient $D$, and $E$/$D$ ratio are available in the Supplementary Material.

We have found that the order of magnitude of the isotropic coupling coefficient $D$ is conserved for most intermonomer configurations available during low-energy NaLi+Na collision.
In the scattering calculations reported below, we neglected the interaction-induced variation of the spin-spin coupling. This means the modification of the spin-spin coupling in a NaLi molecule due to its interaction with a Na atom is omitted, but the geometry dependence of the spin-spin coupling in a NaLi molecule on its orientation within a NaLi+Na complex is included.

\subsection{Geometry and interpolation}

 We use Jacobi coordinates to parameterize the geometry of NaLi+Na complex. The interatomic separation in NaLi is fixed at the vibrationally averaged interatomic distance of 9.139$~a_0$ for the ground state of NaLi ($^3\Sigma^+$)~\cite{gronowski:20}. We perform the computations for more than 20 values of the atom-molecule separation $R$. The cosine of angle ($x_i=\cos(\theta_i), i \in \{0,1,2,..,13,14\}$) are selected from the roots of the 15-th order Legendre polynomial, and are ordered as $x_i>x_{i+1}$. We independently calculated the nonadditive three-body contribution [$V^{3b}(R,\theta)$] and electron spin-spin interaction coupling [$D(R,\theta)$] for each combination of geometric parameters.
For each distance, we interpolate the three-body interaction term and the spin-spin interaction coupling constants by Legendre polynomials \cite{Ogura1920}. We determine Legendre expansion coefficients using the Gauss–Legendre quadrature as
\begin{equation}\label{eq:LEC}
V_L^{(3b)}(R)=(2L+1) \sum_{i=0}^{n-1} w_i P_L(x_i) V^{3b}(R,\theta_i),
\end{equation}
and similar for the spin-spin coupling, and $w_i$ are the quadrature weights.
For use in scattering calculations, the radial dependence of the expansion coefficients is interpolated using the reproducing kernel Hilbert space method \cite{ho:96}.

\subsection{Accuracy \label{sec:accuracy}}

The global minimum of the molecule-atom interaction potential for the quartet state of NaLi+Na with the internuclear distance in NaLi fixed at 9.139$\,a_0$ occurs at $R_e=7.48\,a_0$ and $\theta_e=117.3\,^\text{o}$ with the well depth of $D_e=812.5\,$cm$^{-1}$ (see Fig.~\ref{fig:potentials}). At this equilibrium geometry two-body and three-body interactions contribute 230.7$\,$cm$^{-1}$ and 581.8$\,$cm$^{-1}$ to the binding energy, respectively. The CCSD(T) method reproduced 555.0$\,$cm$^{-1}$ (95\%) of the nonadditive part and the CCSDT correction accounts for 26.8$\,$cm$^{-1}$. The final uncertainty of the calculated nonadditive three-body potential is about 17$\,$cm$^{-1}$ (3$\,\%$) at the global minimum of the potential energy surface and results mainly from the incompleteness of the basis set in the CCSD(T) ($\sim$5$\,$cm$^{-1}$) and CCSDT ($\sim$8$\,$cm$^{-1}$) computations, and the lack of core-electron correlation at the CCSDT level ($\sim$3$\,$cm$^{-1}$). We estimate the magnitude of the neglected contribution of core and core-valence correlation beyond CCSD(T) level by running all-electron CCSDT calculations with the virtual one-electron space truncated by 20$\,$\%~\cite{Kallay2011}. 

Our calculations show that the three-body term is larger than the two-body one and predominantly attractive. Consequently, the intermonomer NaLi distance is shorter than the intramolecular one. The interatomic distances in the complex are closer to the classical turning point of diatomic potentials than to their minima. The repulsive part of the two-body potential is less accurate than its long-range part. This introduces additional uncertainty to the overall potential, which can be as high as 16 cm$^{-1}$ (an additional uncertainty of $\sim$3$\,$cm$^{-1}$ of the three-body interaction), estimated from the difference between an experimental two-body potential for Na$_2$~\cite{Bauer2019} and a highly accurate theoretical potential for NaLi~\cite{gronowski:20}.

The native uncertainty of our state-of-the-art electronic calculation of the three-body term is relatively small, although still too large to predict molecular scattering lengths. Additional uncertainty is introduced by using the rigid rotor approximation to describe NaLi within NaLi+Na. This approximation works very well for van der Waals complexes of deeply-bound molecules, but for high-spin weakly bound alkali-metal systems may result in non-negligible errors~\cite{TomzaPRA13}. Our calculations show that the global equilibrium geometry for NaLi+Na with relaxed distance in NaLi has isosceles triangular symmetry and internuclear distances smaller by around 20\% than in the diatomic molecules due to large three-body forces, the magnitude of which monotonically increases with decreasing atom-molecule separation. 
To reflect the errors related to the difference in three-body interaction at the global minimum and the minimum within the rigid rotor approximation used,
as well as the increased uncertainty of the two-body interactions at these shorter interatomic distances,
we used a more conservative estimate of the uncertainty of the three-body term in the remainder of this paper.

\section{Scattering calculations}

\subsection{Basis set and Hamiltonian}

In our coupled-channels calculations the scattering wavefunction is expanded in a basis of fully coupled channel functions of the form
\begin{align}
|(N L) J (s~s_3) S;& \mathcal{J} \mathcal{M} \rangle = \sum_{M_J, M_S} \langle J M_J S M_S | \mathcal{J} \mathcal{M} \rangle \nonumber \\
\times& |(N L) J M_J\rangle |(s~s_3) S M_S \rangle
\label{eq:basis}
\end{align}
where $\langle j_1 m_1 j_2 m_2 | j m\rangle$ is a Clebsch-Gordan coefficient.
The quantum number $N$ represents the rotational angular momentum of the NaLi molecule,
and $L$ the angular momentum associated with the end-over-end rotation of the atom and molecule about one another,
which are Clebsch-Gordan coupled to a total mechanical angular momentum $J$ with $B$-field projection $M_J$.
Similarly, $s$ denotes the molecular electronic spin resultant from coupling atomic spins $s_1$ and $s_2$ within NaLi,
whereas $s_3$ is the electronic spin of Na,
and $S$ the total electronic spin with $B$-field projection $M_S$.
In the coupled basis, $J$ and $S$ are subsequently coupled to a total angular momentum $\mathcal{J}$ with magnetic-field projection $\mathcal{M} = M_J + M_S$.
Nuclear spin is initially not taken into account, see Sec.~\ref{sec:hf}.
We assume the molecular bond length fixed at the triplet equilibrium position.

The total angular momentum projection along the magnetic field axis $\mathcal{M} = M_J + M_S$ is strictly conserved.
For large enough magnetic field $M_S$ becomes a good quantum number, and therefore also $M_J=\mathcal{M}-M_S$ is good.
The Na atomic spin is $s_3=1/2$ throughout.
Due to the large singlet-triplet splitting in the NaLi molecule, $s=0$ or 1 is also a good quantum number.
For a separated atom and molecule, $m_s$ and $m_{s_3}$ would separately become good quantum numbers,
but at chemically relevant distances the exchange splitting between the doublet and quartet interaction potentials is dominant such that $S=1/2$ and $3/2$ are good quantum numbers.
Hence, we can effectively consider each $|S M_S\rangle$ state separately,
with only perturbatively weak couplings between them.
For each of these spin channels, there are strong and anisotropic interactions that couple different $N$ and $L$ channels,
but conserve $J$ and $M_J$.
The initial channel corresponds to $s$-wave collisions in the spin-stretched rotational ground state,
$|(N L) J M_J\rangle |S M_S \rangle = |(0 0) 0 0\rangle|3/2\ 3/2\rangle$ for the low-field seeking upper spin-stretched state,
and $|(0 0) 0 0\rangle|3/2,\ -3/2\rangle$ for the high-field seeking lower spin-stretched state, respectively.

In our coupled-channels calculations we include the electronic interaction described in Sec.~\ref{sec:interaction},
\begin{align}
\hat{V}(R,\theta) &= \sum_L P_L(\cos\theta) \times \Big[ V^{(0)}_L(R) + \hat{s}_1\cdot\hat{s}_2 V^{(12)}_L(R)\nonumber \\
+& \hat{s}_1\cdot\hat{s}_3 V^{(13)}_L(R) + \hat{s}_2\cdot\hat{s}_3 V^{(23)}_L(R) + V_L^{(3b)}(R) \Big],
\end{align}
the Zeeman interaction with the magnetic field,
\begin{align}
\hat{H}_\mathrm{Zeeman} = \mu_B g_e B (\hat{s}_{1,z}+\hat{s}_{2,z}+\hat{s}_{3,z}),
\end{align}
the magnetic dipole-dipole interaction
\begin{align}
\hat{V}_\mathrm{magn. dip} =& -\sqrt{30} \frac{(\mu_B g_e \alpha)^2}{R^{3}} \nonumber \\
&\times \left[\left[\hat{s} \otimes\hat{s}_3 \right]^{(2)} \otimes C^{(2)}(\hat{R}) \right]^{(0)}_0,
\end{align}
and the spin-rotation 
\begin{align}
\hat{H}_\mathrm{spin-rotation} = \gamma_s \hat{N} \cdot \hat{s},
\end{align}
and spin-spin couplings
\begin{align}
\hat{H}_\mathrm{spin-spin} = \lambda_s \sqrt{30}/3 \left[\left[\hat{s} \otimes\hat{s}\right]^{(2)} \otimes C^{(2)}(\hat{r}_\mathrm{NaLi}) \right]^{(0)}_0,
\end{align}
where $[\hat{A} \otimes \hat{B}]^{(k)}_q$ indicates a tensor product and $C^{(2)}(\hat{R})$ indicates spherical harmonics, see the Appendix.
The spin-spin coupling parameter $\lambda_s = 0.0189$~cm$^{-1}$ is computed here,
whereas for the spin-rotation coupling we use the $\gamma_s=0.005$~cm$^{-1}$ upper limit estimated in Ref.~\cite{gronowski:20}.
The reader is referred to the Appendix for a full description of the Hamiltonian and for matrix elements in the channel basis.

The spin-rotation coupling $\gamma_s \hat{N} \cdot \hat{s}$ and spin-spin coupling $\lambda_s \left[ \left(\hat{s} \cdot \hat{r}_\mathrm{NaLi}\right)\left(\hat{s} \cdot \hat{r}_\mathrm{NaLi}\right) - \frac{1}{3} \hat{s}^2\right]$ play an important role here as these are spin-dependent couplings that are not diagonal in $S$ and $M_S$.
Losses by Zeeman relaxation or transitions to the chemically reactive doublet potential must involve these couplings.
It is worth noting that if the electronic interaction were isotropic, \emph{i.e.}, independent of the relative orientation of the atom and molecule,
these spin-dependent couplings would not be enough to lead to Zeeman relaxation.
However, after accounting for the spin-rotation and spin-spin coupling,
molecular eigenstates in different Zeeman levels have different rotational-state decompositions, and hence are coupled by the anisotropic part of the electronic interaction.
That is, physically, the strong anisotropic electronic interaction can reorient the molecule which effectively flips the electronic spin because the spin is coupled to the molecular axis by spin-rotation and spin-spin coupling.
The role of anisotropic interactions implies that transitions occur at short range and cannot be described by simpler models based on isotropic long-range $R^{-6}$ interactions alone~\cite{wang:21c}.

\begin{figure}
\begin{center}
\includegraphics[width=0.9\columnwidth]{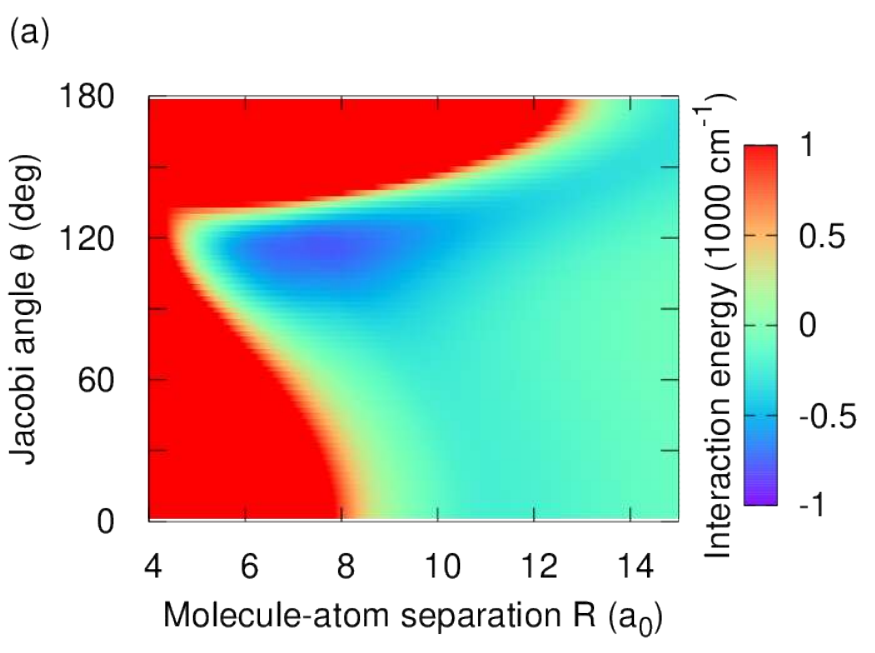}
\includegraphics[width=0.9\columnwidth]{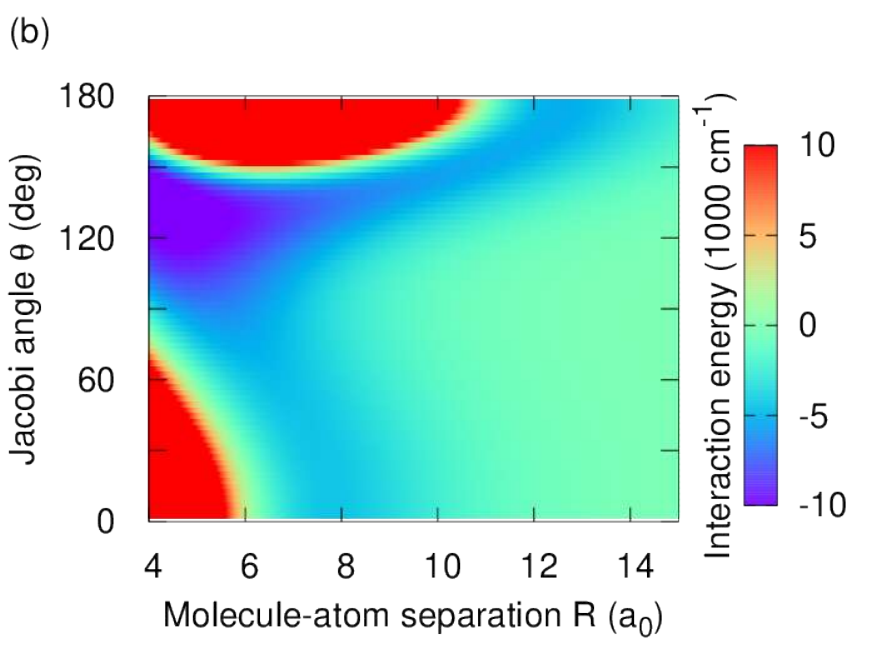}
\caption{ \label{fig:potentials}
{\bf Interaction potentials for the NaLi+Na collision complex.}  The potentials are shown as a function of $R$, the center of mass distance between the atom and molecule,
and $\theta$ the Jacobi angle between the orientation of the molecule and the approach of the atom, with $\theta=0$ corresponding to Na-NaLi and $\theta=180^\circ$ to Na-LiNa.
The molecular bond length is fixed at the triplet equilibrium bond length $r=8.8~a_0$.
{\bf (a)} high-spin quartet $S=3/2$ potential.
{\bf (b)} low-spin doublet $S=1/2$ potential.
In principle there are two doublet potentials that can have an avoided crossing.
What is shown here is the doublet potential for a pure triplet NaLi molecule, $s=1$.
}
\end{center}
\end{figure}

After having described the atom-molecule interactions, approximately good quantum numbers, and the critical role of various coupling mechanisms,
we continue by performing full quantum mechanical coupled-channels calculations.
For computational tractability we will start out ignoring hyperfine and vibrational degrees of freedom.
By ignoring the vibrational coordinate of the molecule, and fixing the molecular bond length to the triplet NaLi equilibrium bond length, our model cannot directly describe chemical reactions.
The only energetically accessible products are to form singlet NaLi or Na$_2$ molecules. Chemical reactions will occur on the low-spin potential.  
In our coupled-channels calculations, we model these by imposing an absorbing boundary condition at $R_\mathrm{min}=4.5~a_0$, which can be reached on the low-spin potential, but not on the high-spin potential which is highly repulsive at these short distances, see Fig.~\ref{fig:potentials}.
This choice seems arbitrary but it does not affect the results as long as the boundary condition is imposed in a region where the high-spin potential is highly repulsive and simultaneously the low-spin potential is strongly attractive.
Any flux that reaches this region in the low-spin state will continue classically to smaller $R$,
such that it does not matter where exactly in this region one matches to the absorbing boundary condition.
We have confirmed this numerically for $R_\mathrm{min}$ between 4 and 4.5~$a_0$.

\subsection{Cross sections and rate coefficients}

We solve the coupled-channels equations numerically using the renormalized Numerov propagator.
Using the method of \cite{janssen:13}, and described in more detail in Ref.~\cite{karman:23},
we match to reactive boundary conditions at short range and the usual scattering boundary conditions at long range.
Again, the short-range boundary condition is imposed at $R_\mathrm{min}=4.5~a_0$,
a separation that can be reached only on the low-spin potential and effectively models chemical reactions.
This procedure yields an ``inelastic'' $S$-matrix, $\bm{S}^{(\mathrm{LR})}$ and a ``reactive'' $S$-matrix $\bm{S}^{(\mathrm{SR})}$.
The elements of the inelastic $S$-matrix, ${S}^{(\mathrm{LR})}_{f,L'; i,L}$, describe the amplitudes for scattering from an initial state $i$ and partial wave $L$, to a final state $f$ and partial wave $L'$.
The elements of the reactive $S$-matrix, ${S}^{(\mathrm{SR})}_{r; i,L}$ describe scattering from an initial state $i$ and partial wave $L$ into a reactive channel $r$ at short range.
The reactive channels are determined by diagonalizing the Hamiltonian excluding radial kinetic energy at the short-range matching point.
From the $S$-matrices one can determine the elastic cross section
\begin{align}
    \sigma^\mathrm{elastic} = \frac{\pi}{k^2} \sum_{L,L'} \left| \delta_{L,L'}-S^{(\mathrm{LR})}_{i,L'; i,L} \right|^2,
\end{align}
where $i$ is the initial state and $k = \hbar^{-1} \sqrt{2\mu E}$ is the intial wavenumber.
From the $S$ matrices one can also determine the inelastic rate coefficient
\begin{align}
    k^{(\mathrm{inelastic})} = \frac{\pi}{\mu k} \sum_{L,L',f\neq i} \left| S^{(\mathrm{LR})}_{f,L'; i,L} \right|^2,
\end{align}
the rate coefficient for loss at short range
\begin{align}
    k^{(\mathrm{short})} = \frac{\pi}{\mu k} \sum_{L,r} \left| S^{(\mathrm{SR})}_{r; i,L} \right|^2,
\end{align}
and we define a total loss rate coefficient, $k^{(\mathrm{loss})} = k^{(\mathrm{inelastic})} + k^{(\mathrm{short})}$.
Both the elastic cross section and the rate coefficients are energy independent for energies well below the van der Waals energy which is in the order of 500~$\mu$K for NaLi+Na.
We computed the cross sections and loss rate coefficients at a single collision energy of 4~$\mu$K.

\subsection{Convergence}

\begin{figure}
\begin{center}
\includegraphics[width=0.9\columnwidth]{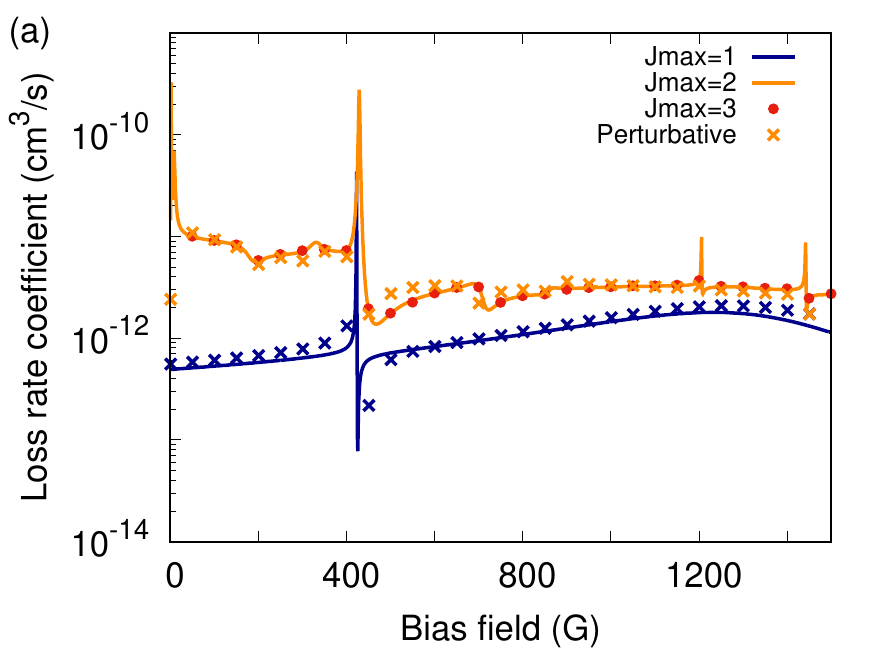}
\includegraphics[width=0.9\columnwidth]{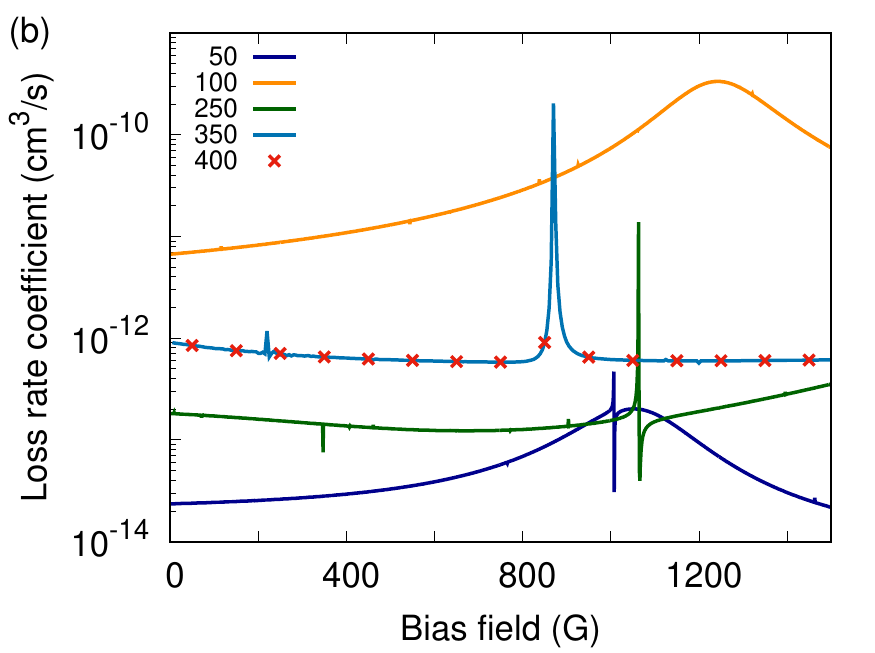}
\includegraphics[width=0.9\columnwidth]{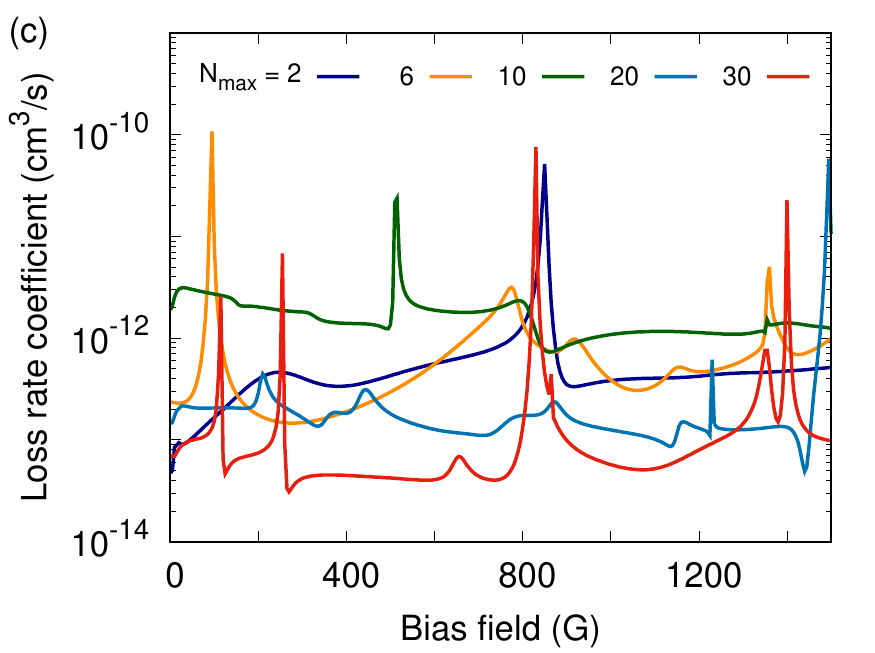}
\caption{ \label{fig:convergence}
{\bf Convergence of the coupled-channels calculations with truncation of the basis set.}
{\bf (a)} Calculation with fixed $N_\mathrm{max}=20$.
Cross sections converge with $J_\mathrm{max}=2$.
When the basis is truncated at $J_\mathrm{max}=1$, the cross sections scale quadratically with the spin-rotation coupling constant $\gamma_s$.
This is demonstrated by agreement with the crosses which show one-quarter of the cross section obtained by scaling $\gamma_s$ up by a factor of two.
When the basis is truncated at $J_\mathrm{max}=2$, spin-spin coupling contributes and the cross section scales perturbatively with both coupling constants (crosses).
Spin-spin coupling is typically dominant, but not by a large factor so both mechanisms contribute.
Magnetic dipole-dipole coupling does not play an important role.
{\bf (b)} Calculation with fixed $J_\mathrm{max}=1$
The scattering cross section can be converged with $N_\mathrm{max}$, but requires an impractically large basis set.
{\bf (c)} Calculation with fixed $J_\mathrm{max}=2$
The scattering cross section is not converged with $N_\mathrm{max}$, but the density and typical width of resonances are not strongly dependent on $N_\mathrm{max}$.
}
\end{center}
\end{figure}

First we consider convergence of the calculation with $J_\mathrm{max}$, the highest value of $J$ included in the basis set.
As explained above, we can essentially consider each spin channel $|S M_S\rangle$ independently with only perturbative couplings between them,
and within each spin level, we can consider $J$ and $M_J$ to be good quantum numbers.
Spin-rotation coupling has the selection rules $\Delta J = 0 \rightarrow 1$ and $|\Delta M_S| \le 1$.
Spin-spin coupling has the selection rules $\Delta J = 0 \rightarrow 2$ and $|\Delta M_S| \le 2$.
Hence, if these spin-dependent couplings act only perturbatively we expect cross sections do not change by including functions with $J=3$ or higher,
and that cross sections scale quadratically in the spin-rotation and spin-spin coupling strengths.
This is exactly what we observe in Fig.~\ref{fig:convergence}(a).
Inclusion of functions with $J=3$ does not affect the cross sections (dots),
and the cross sections scale perturbatively with the spin-rotation and spin-spin coupling strengths (crosses).
The contribution of the magnetic dipolar interaction between the atomic and molecular magnetic moment is far smaller and we have discussed in less detail for this reason, although it is included in the calculation.
Finally we note that the mechanism responsible for the loss rates requires anisotropic electronic interactions,
but the anisotropy is not perturbatively weak.

The convergence of the calculation with $N_\mathrm{max}$ for fixed $J_\mathrm{max}=1$ is shown in Fig.~\ref{fig:convergence}(b).
The expectation is that the calculation converges when all locally open channels are included,
\emph{i.e.}, when the excitation energy of channels that are excluded are all higher than the depth of the interaction potential.
\cite{morita:18,morita:19,morita:20}
If we estimate the required $N_\mathrm{max}$ based on the depth of the spin-stretch potential of 800~cm$^{-1}$, one would expect the calculation to converge with $N_\mathrm{max} \approx 70$.
Instead, we see the calculation converges with much higher $N_\mathrm{max}=350$.
The reason for this appears to be that these higher rotational channels continue to contribute locally open channels in the low-spin state that affect the background scattering length.

As we have seen, converging the scattering calculation requires including also functions with $J=2$.
Scattering calculations for basis sets truncated at $N_\mathrm{max}=350$ and $J_\mathrm{max}=2$ become prohibitively computer intensive.
Fortunately, the typical number of resonances appears to converge more rapidly with $N_\mathrm{max}$, and can be predicted with much lower truncation of $N$.
Figure~\ref{fig:convergence}(c) shows loss rates computed with various $N_\mathrm{max}$ up to 30 for fixed $J_\mathrm{max}=2$.
We will come back to convergence of the typical density of resonances after discussing the dependence on the interaction potential.

\begin{figure}
\begin{center}
\includegraphics[width=0.9\columnwidth]{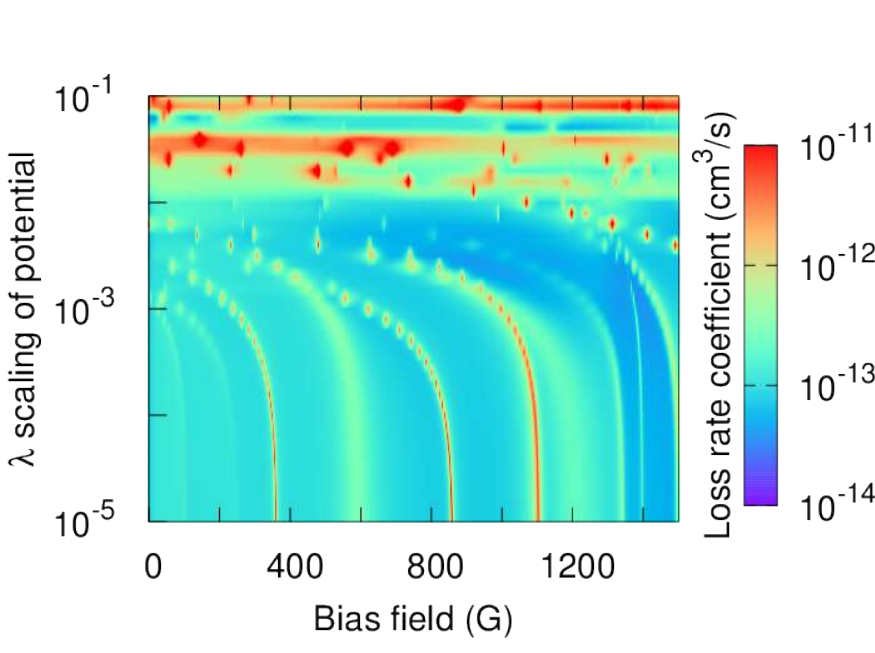}
\caption{ \label{fig:lambda}
{\bf Dependence of loss rates on scaling of three-body interactions.} Loss rates as a function of $B$-field and scaling of the three-body interaction by a factor $1+\lambda$ for $N_\mathrm{max}=20$,
shown for the upper stretched state.
This shows that due to the uncertainty of $\lambda$ of at least several percent, the background loss and resonance positions are undetermined.
Several $B$-independent resonances are observed, where the $\lambda$-scaling tunes the initial spin-stretched potential such that it supports a resonance, i.e., a bound state near zero energy.
Hence, prediction of resonance positions requires knowledge of the interaction potentials to an accuracy that cannot realistically be achieved by \emph{ab initio} calculations.
}
\end{center}
\end{figure}

\begin{figure}
\begin{center}
\includegraphics[width=0.9\columnwidth]{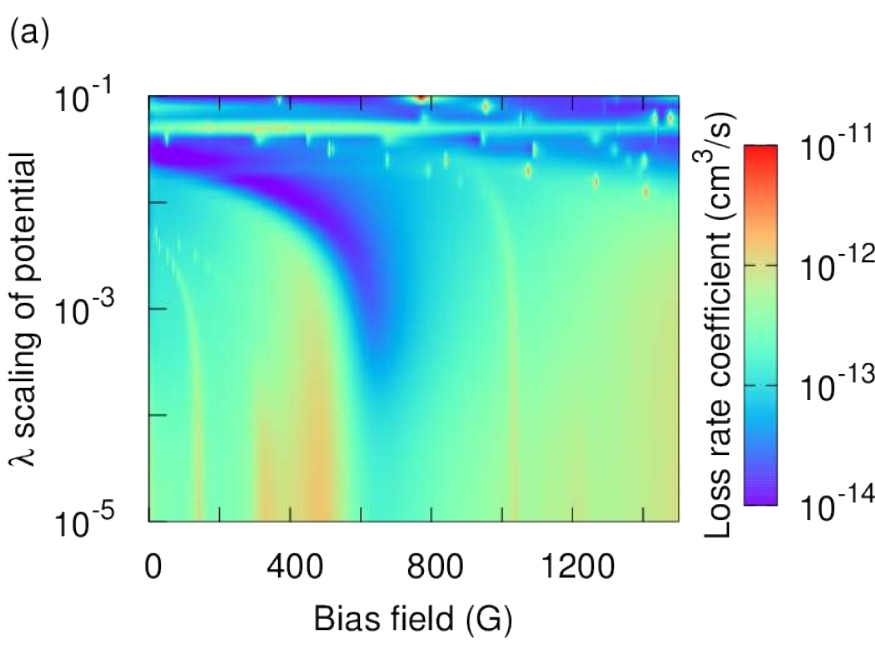}
\includegraphics[width=0.9\columnwidth]{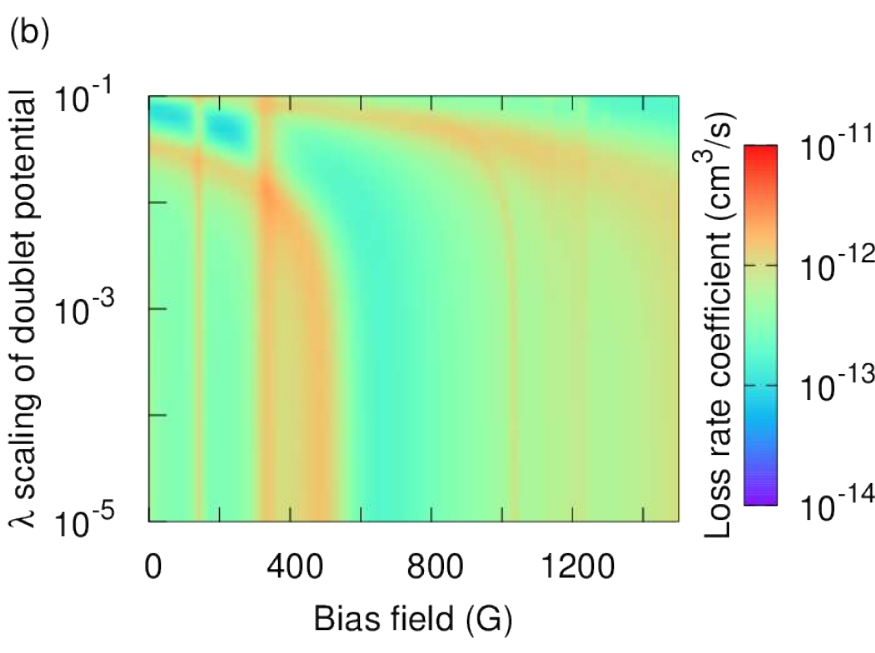}
\caption{ \label{fig:lambda6}
{\bf Spin-dependence of three-body interactions.}  Loss rates are shown as a function of $B$-field and scalings of the interactions potential.
Calculations are done with $N_\mathrm{max}=6$, which artificially reduced the number of resonances compared to Fig.~\ref{fig:lambda}.
{\bf (a)} The spin-independent three-body interaction is scaled by $1+\lambda$.
In this case, all resonance positions depend on the scaling of the potential,
the background loss rate varies strongly,
and new $B$-independent resonances appear by scaling the initial-state potential such that supports a bound state near zero energy.
{\bf (b)} The spin splitting is scaled by $1+\lambda$, while the spin-stretched potential is kept fixed. 
Therefore, only the doublet potential is varied.
The resonances near 150, 350, 1150, and 1250~G are independent of the scaling of the low-spin potentials, i.e., they are completely supported by the non-reactive spin-stretched potential.
The feature just above 1000~G has a weak dependence on the scaling of the doublet potentials,
and several broader features such as that around 500~G have a stronger dependence.
}
\end{center}
\end{figure}

Figure~\ref{fig:lambda} shows the sensitivity of the scattering rates to the interaction potential, parameterized by $\lambda$.
Here, we scale by a factor $1+\lambda$ the non-additive three-body part of the interaction potential, \emph{i.e.}, the part that is computed \emph{ab initio} and is uncertain up to an estimated 3~\% within the rigid rotor approximation for NaLi, and considerably larger when this approximation is relaxed, see Sec.~\ref{sec:accuracy}.
Figure~\ref{fig:lambda} is computed with $J_\mathrm{max}=2$ and $N_\mathrm{max}=20$.
By modifying the three-body interaction by less than $10^{-4}$, we find that the resonances are unaffected.
By modifying the potential by about $10^{-3}$, we find that the resonances start to shift such that realistically their positions are completely undetermined.
When the scaling of the potential is at the level of several percents, we find that horizontal $B$-field independent resonances appear.
This occurs when the potential is scaled to support a resonance near zero collision energy for the \emph{initial} Zeeman level, which is therefore not tuned by the magnetic field.
Therefore, with realistic uncertainties in the three-body part of the interaction and the rigid rotor approximation, both the position of magnetically tunable resonances and the background scattering length are completely undetermined but we can still draw probabilistic conclusions about a typical range of values for the scattering length.

Figure~\ref{fig:lambda6} shows a similar ``$\lambda$-scan'', but now for $N_\mathrm{max}=6$ which artificially reduces the density of resonances somewhat and produces a more sparse figure.
In panel~\ref{fig:lambda6}(a) we have scaled the non-additive three-body part of the interaction potential for both the high-spin and the low-spin potential, as before.
In panel~\ref{fig:lambda6}(b), by contrast, we have assumed this uncertainty is entirely in the low-spin potential, and we have kept the high-spin potential constant.
In this case, we no longer observe horizontal $B$-field resonances as these are supported by the initial top Zeeman energy level, which is a high-spin state.
We find that several of the resonances are now completely independent of the scaling of the low-spin potential up to $\lambda=0.1$.
This indicates the resonances are supported by the non-reactive high-spin potential.
When the scaling of the low-spin potential reaches the percent level,
coupling to the low-spin state starts to affect the collision rate, especially for the broader features.

The analysis above shows that the resonances are supported by the high-spin potential,
and that their positions are sensitive to 0.1~\% uncertainty in the non-additive three-body part of the interaction potential.
This means that the \emph{ab initio} prediction of the resonance positions is beyond the capability of state-of-the-art theory.
In addition to highly accurate atom-molecule interactions,
the quantitative prediction of resonance positions would require converged coupled-channels calculations that also fully account for hyperfine, molecular vibrations, and chemical reactions.
This is not attempted here, and instead we interpret only the typical number of resonances, their widths, and coupling mechanisms.

\subsection{Background elastic and inelastic scattering}

Next we consider interaction potentials obtained by scaling the non-additive three-body interactions by $1+\lambda$ with $\lambda$ between $-0.1$ and $+0.1$, on the order of the uncertainty in the interaction potential.
We consider each of these Hamiltonians statistically independent realizations of the physical NaLi+Na system.
By performing scattering calculations for these different realizations, we gather statistics about the number of resonances and their widths.

\begin{figure}
\begin{center}
\includegraphics[width=0.9\columnwidth]{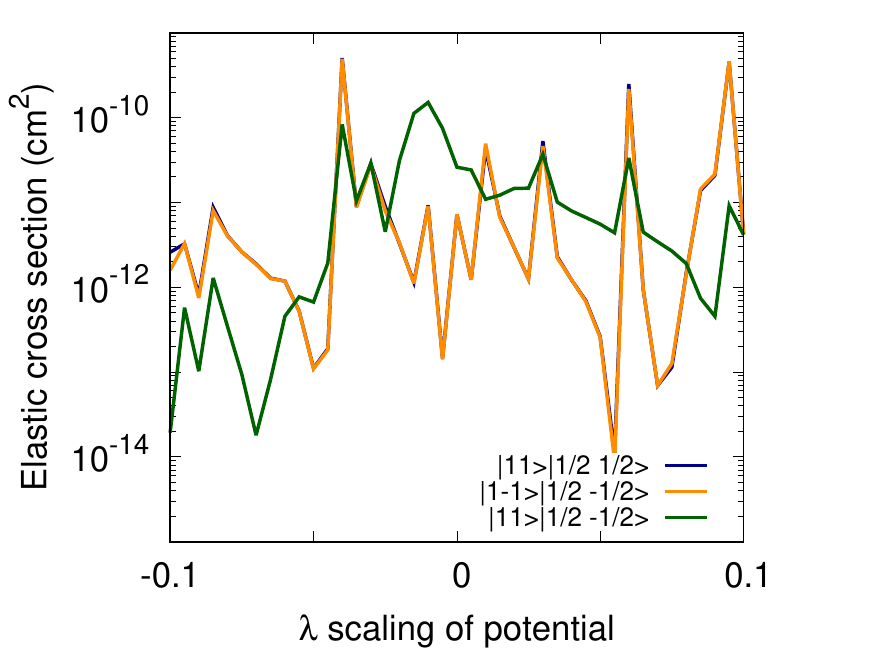}
\includegraphics[width=0.9\columnwidth]{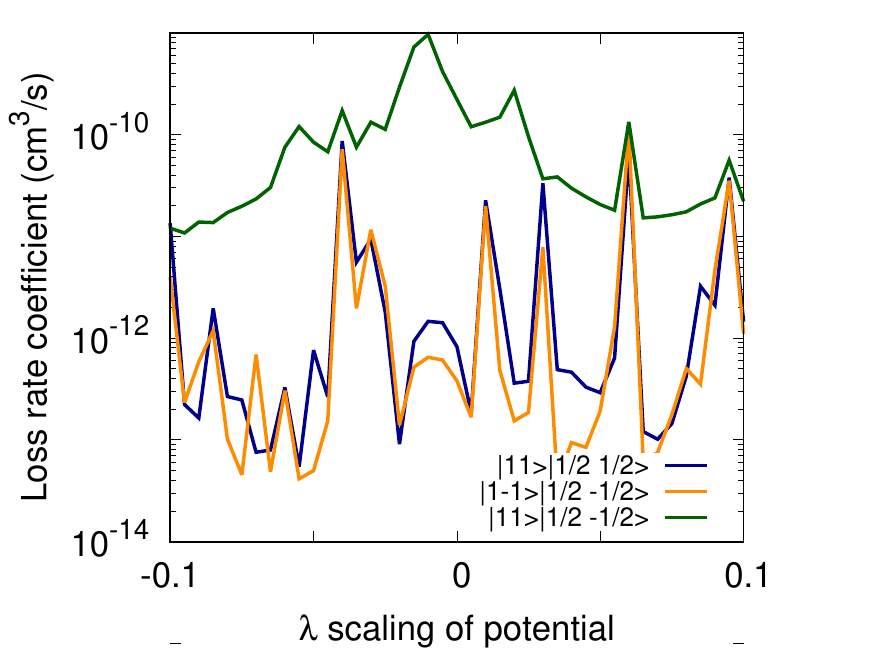}
\end{center}
\caption{ \label{fig:lambda0B}
{\bf Cross sections and rates as a function of $\lambda$ scaling.}  {\bf (a)} The elastic collision cross section obtained by scaling the non-additive three-body interaction by $1+\lambda$ at a fixed magnetic field $B=1500$~G.
Different colors correspond to different spin states, $|s m_{s}\rangle|s_3 m_{s_3}\rangle$, as indicated.
{\bf (b)}
Loss rate coefficients as a function $\lambda$ scaling for fixed $B=1500~$G.
}
\end{figure}

\begin{figure}
\begin{center}
\includegraphics[width=0.9\columnwidth]{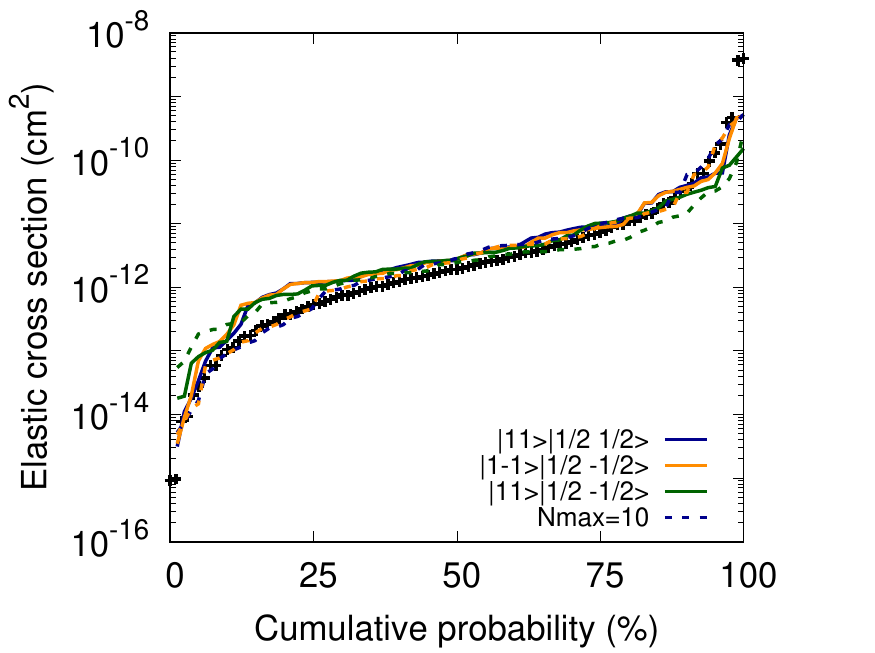}
\includegraphics[width=0.9\columnwidth]{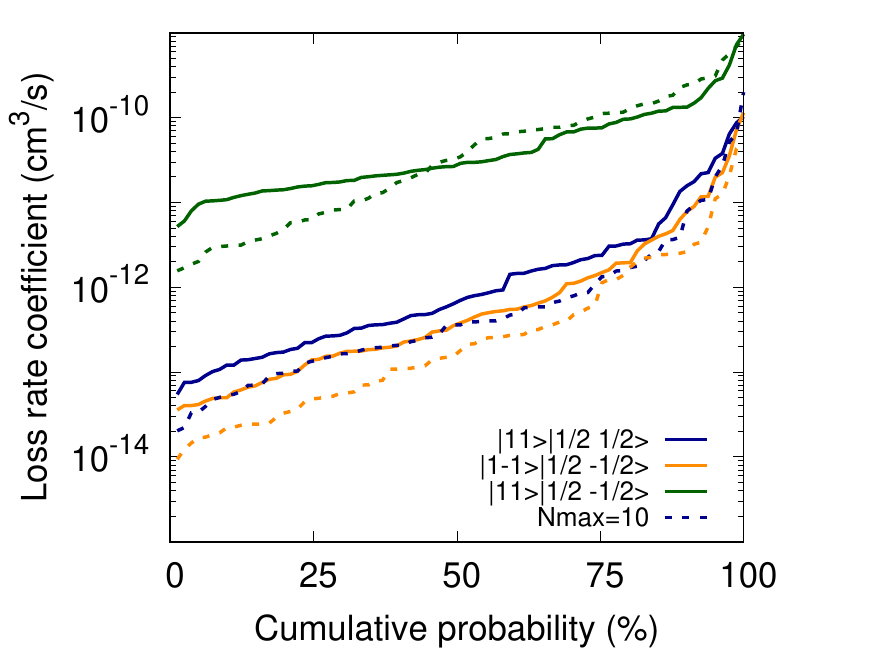}
\includegraphics[width=0.9\columnwidth]{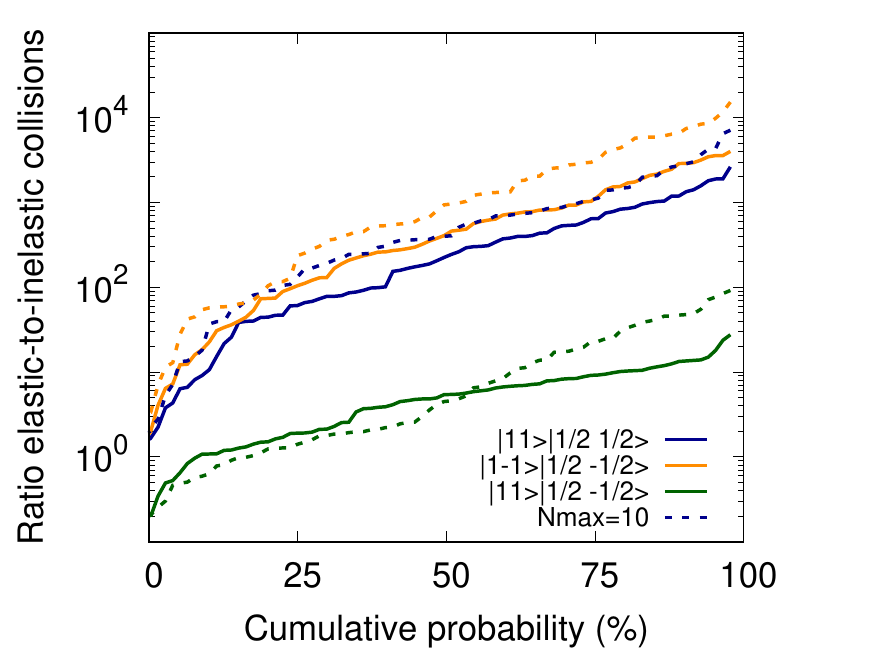}
\end{center}
\caption{ \label{fig:lambda1B}
{\bf Probability distribution of collision rates.}  
Cumulative probability distribution obtained by sorting cross sections and rate coefficients for different $\lambda$ scaling at fixed $B=1500$~G \cite{morita:19}.
Different colors correspond to different states, $|s m_{s}\rangle|s_3 m_{s_3}\rangle$, as indicated.
{\bf (a)} Distribution of the elastic cross section,
{\bf (b)} loss rate coefficient,
and {\bf (c)} the ratio of elastic-to-inelastic collisions.
Solid (dashed) lines correspond to $N_\mathrm{max}=20$ ($10$).
In panel(a) the black markers indicated the expected distribution of elastic cross sections for a van der Waals potential~\cite{gribakin:93}.
}
\end{figure}

We first consider the nonresonant background.
Figure~\ref{fig:lambda0B} shows the dependence of the elastic cross section and loss rate coefficient on scaling of the non-additive three-body interaction by $1+\lambda$ at a fixed magnetic field $B=1500$~G, $N_\mathrm{max}=20$.
These calculations are performed for three initial states;
the top stretched state where the molecular electronic spin projection $m_s=1$ and the atomic electronic spin projection $m_3=1/2$,
the bottom stretched state where $m_s=-1$ and $m_3=-1/2$,
and the non-stretched state $m_s=1$, $m_3=-1/2$.
The elastic cross sections are nearly identical in the two stretched states,
and different but of the same order in the non-stretched initial state.
For the loss rate coefficient, the \emph{typical} behavior is also that the cross sections are similar between the two stretched states,
while the differences at fixed $\lambda$ can be as large as an order of magnitude.
For the non-stretched initial state the loss rate coefficient is significantly larger.
This can be understood as follows:  For non-stretched states the collision can directly proceed on the reactive potential and lead to loss.
For stretched states, the initial potential is nonreactive such that loss processes require a spin flip which is  unlikely since it is perturbative in the weak spin-spin and spin-rotation coupling, as we have seen above.

In Figure~\ref{fig:lambda1B} we show again the effect of scaling of the non-additive three-body interaction by $1+\lambda$,
but ordered by increasing cross section and loss rate.
We interpret the resulting horizontal axis as the cumulative probability to obtain a specific cross section or rate coefficient \cite{morita:19}.
Panel~\ref{fig:lambda1B}(a) shows that elastic cross sections,
for which all cumulative probability distributions are similar and agree closely with the expected distribution for a $R^{-6}$ potential \cite{gribakin:93}, shown as the black markers.
This distribution is characterized by a mean scattering length $\bar{a} = 0.47799 (2\mu C_6/\hbar)^{1/4} \approx 52~a_0$.
The experimentally measured scattering length is 260~$a_0$,
the corresponding elastic cross section is larger than one might expect. 

Panel~\ref{fig:lambda1B}(b) shows the cumulative probability distribution of the loss rate coefficient.
The two spin-stretched states show a similar distribution,
whereas the typical loss rate is much higher in the non-stretched state, as observed before.
Similar probability distributions are obtained with $N_\mathrm{max}=10$ and $20$. 
In the spin-stretched states, the background loss rate is likely to lie between $10^{-13}$ and a few $10^{-12}$ cm$^3$/s.
Experimentally, a typical background loss rate of $5\times 10^{-12}$ cm$^3$/s is observed,
which on the higher end of this distribution function, but not in disagreement with it.

Finally, in panel~\ref{fig:lambda1B}(c), we consider the background ratio of elastic-to-inelastic collisions, $\gamma$. 
We compute the ratio of elastic-to-inelastic collisions as
\begin{align}
\gamma = \frac{\sigma^{(\mathrm{elastic})} \langle v \rangle}{k^{(\mathrm{loss})}},
\end{align}
where $\langle v\rangle$ is the thermally averaged speed at $T=2~\mu$K \cite{son:20},
and we take for the elastic cross section $4\pi a^2$ where $a=260~a_0$ is the measured elastic cross section.
The resulting ratio of elastic-to-inelastic collisions is likely between 100 and 1000 in the spin-stretched states,
in agreement with the experimental value of 300 \cite{son:20}.

\begin{figure*}
\begin{center}
\includegraphics[width=0.475\textwidth]{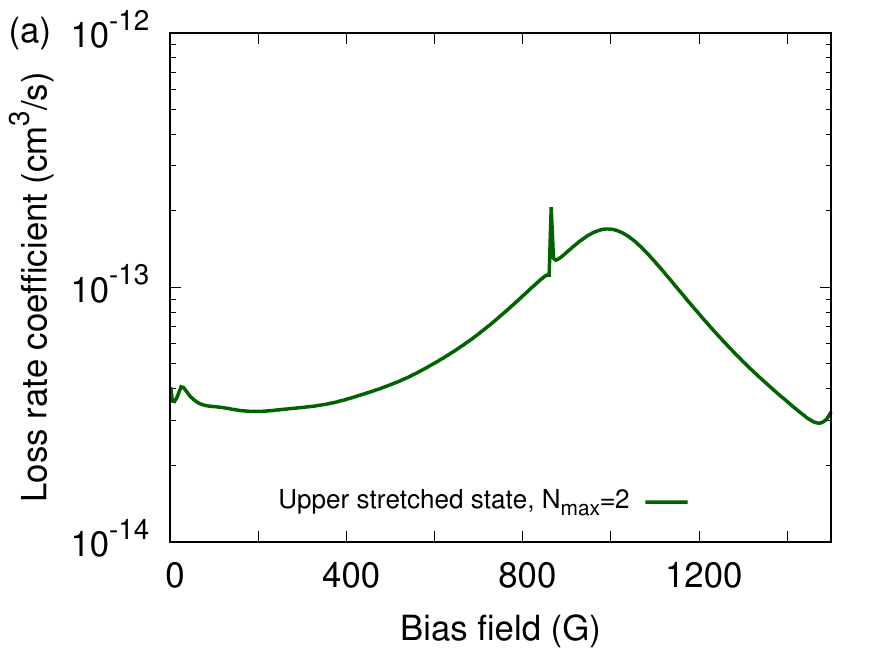}
\includegraphics[width=0.475\textwidth]{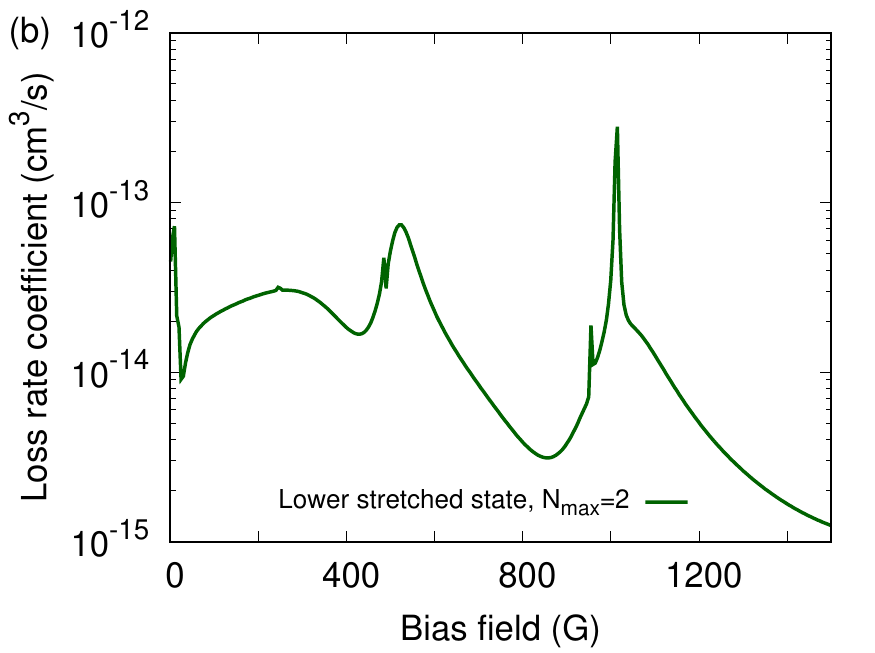}
\includegraphics[width=0.475\textwidth]{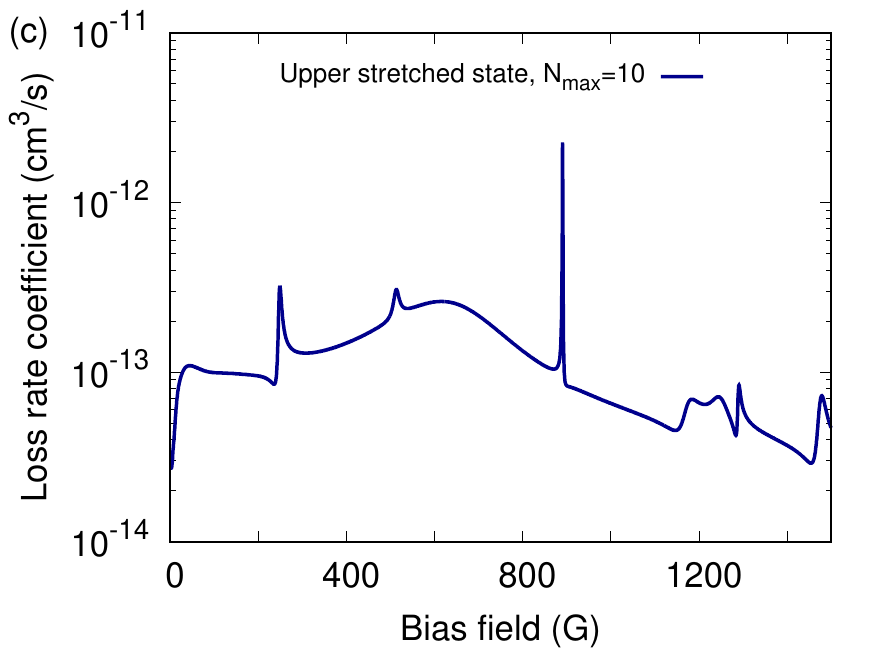}
\includegraphics[width=0.475\textwidth]{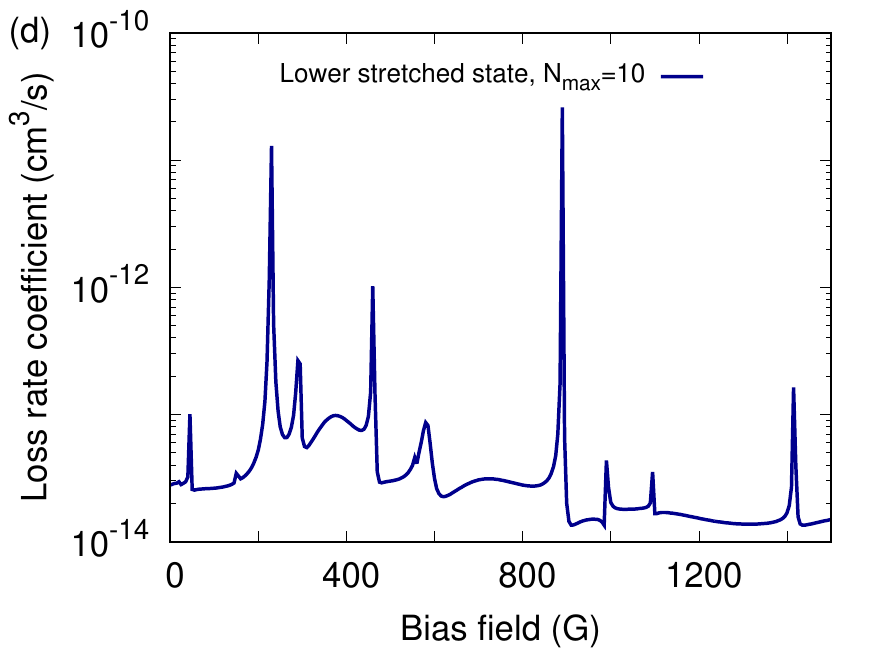}
\includegraphics[width=0.475\textwidth]{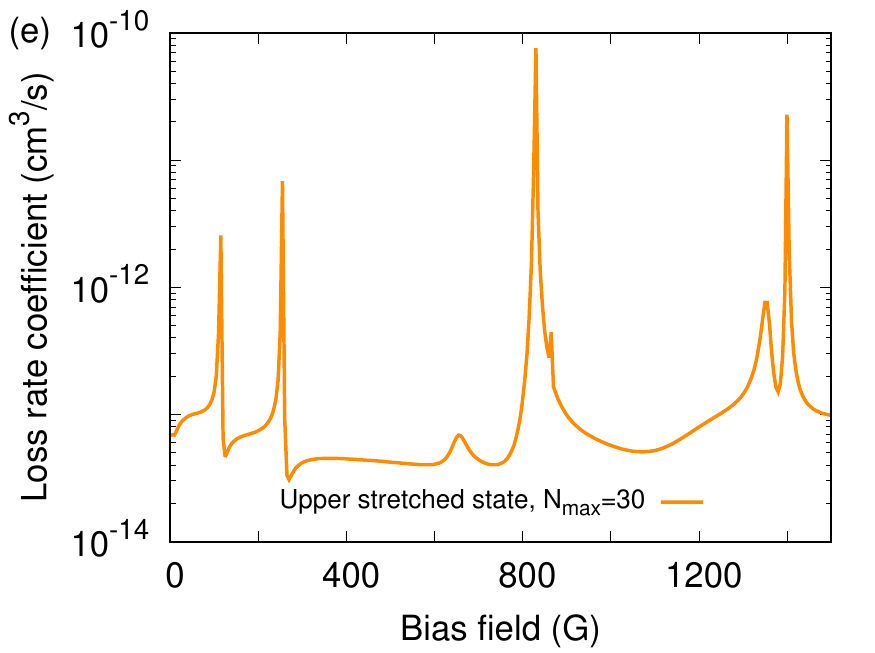}
\includegraphics[width=0.475\textwidth]{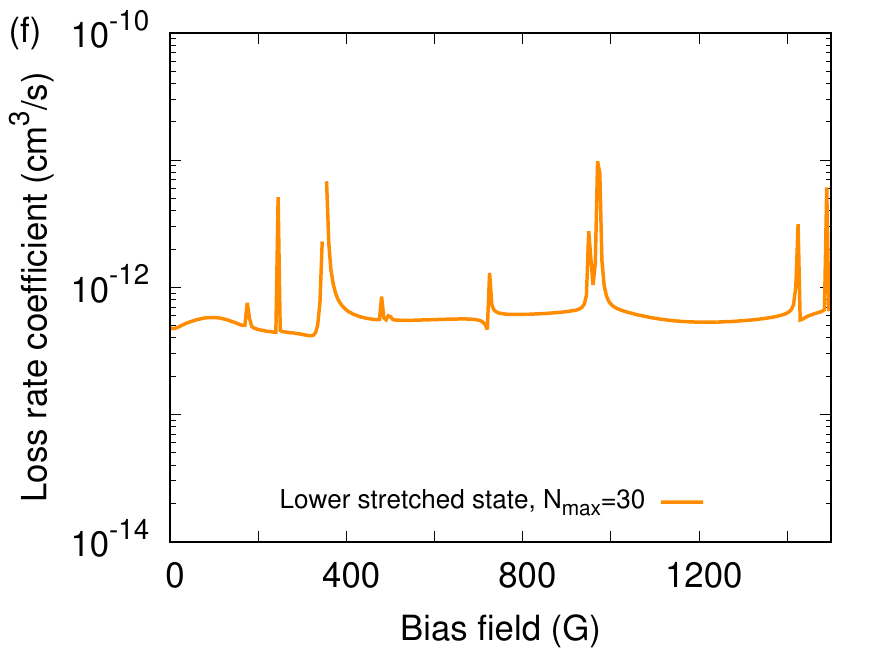}
\caption{ \label{fig:Nconv}
{\bf Representative magnetic field scans for different basis set truncation.}  
The figure shows representative magnetic field scans of the collisional loss rate coefficient. Different panels correspond to truncation of the basis set at $N_\mathrm{max}=2,$ 10, 30, as indicated. The left (right) hand column shows results for the upper (lower) spin-stretched state.
}
\end{center}
\end{figure*}

\subsection{Density of resonances}

Next, we compare the typical number of resonances to the density of states of the spin-stretched NaLi+Na collision complex.
Figure~\ref{fig:Nconv} shows typical magnetic field scans for $N_\mathrm{max}=2$, 10, and 30 for both the top and bottom spin-stretched states.
The density of resonances increases with $N_\mathrm{max}$ and is higher for the lower spin-stretched state than it is for the upper one.

In order to explain the observed number of resonances, we calculate their density of states.
Again, the resonances are supported by the spin-stretched interaction potential,
and the magnetically tunable resonances correspond to Zeeman sub-states that are different from the initial state.
This means each resonance can be assigned total electron spin $S$, $M_S$ quantum numbers.
Spin is coupled to the spatial degrees of freedom perturbatively through spin-spin and spin-rotation coupling.
This means we can assign each state total mechanical angular momentum $J$, $M_J$,
and the spin-rotation and spin-spin selection rules tell us which values of these nearly-good quantum numbers contribute.
Spin-rotation coupling is rank-1 in the spin and spatial degrees of freedom and couples to states with $J=1$ and $|\Delta M_S| \le 1$.
The density of such states is determined by the $J=1$ density of states on the spin-stretched potential.
The total number of bound states below energy $E$ can be computed quasiclassically using a phase-space integral \cite{christianen:19b}.
We determine the total number of resonances expected as the total number of bound states less than the Zeeman energy shift for $\Delta M_S=1$ below threshold.
Spin-spin coupling is rank-2 in the spin and spatial degrees of freedom and couples to $J=2$ and $|\Delta M_S|\le 2$ states.
We similarly determine the number of resonances by computing the number of $J=2$ states bound by less than the Zeeman shifts for $\Delta M_S=1$ and $2$.

\begin{figure}
\begin{center}
\includegraphics[width=0.475\textwidth]{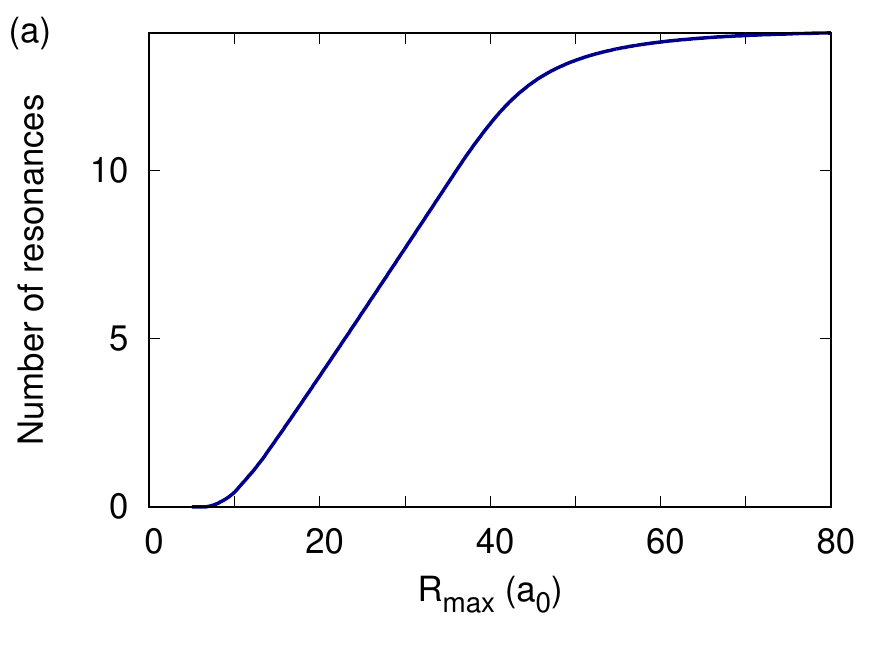}
\includegraphics[width=0.475\textwidth]{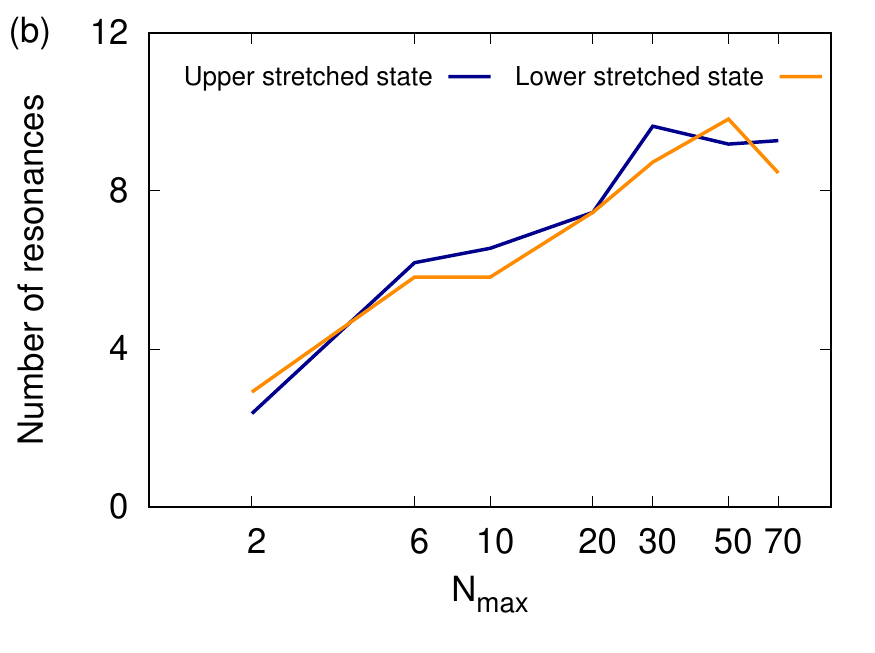}
\caption{ \label{fig:DOS}
{\bf Number of resonances. }  The plot shows the expected number of resonances between 0 and 1500~G from ($\mathbf{a}$) quasiclassical phase-integrals Eqs.~\eqref{eq:rho3d} and \eqref{eq:rho2d} as a function of the maximum molecule-atom center-of-mass separation $R_\mathrm{max}$ up to which the integrals are evaluated and ($\mathbf{b}$) quantum mechanical calculations of bound states on each adiabatic potential energy curve, as a function of $N_\mathrm{max}$ that truncates the channel basis set.
}
\end{center}
\end{figure}

We computed the total numbers of bound states using the phase-space integrals of Ref.~\cite{christianen:19b},
\begin{align}
N^{(\mathrm{3D})} = (2J+1) \frac{g_\mathrm{parity} 8\pi\sqrt{2} m_\mathrm{Na}^2m_\mathrm{Li}}{3 h^3 (2m_\mathrm{Na}+m_\mathrm{Li}) } \nonumber \\
\times \iiint \frac{rR}{\sqrt{\mu R^2 + \mu_\mathrm{NaLi} r^2}} \left[E-V(\bm{q})\right]^{3/2}\ dr\ dR\ d\theta
\label{eq:rho3d}
\end{align}
as well as the number of bound states for the NaLi vibrational coordinate $r=r_e$ fixed at the equilibrium distance
\begin{align}
N^{(\mathrm{2D})} = (2J+1) \frac{g_\mathrm{parity} 2\pi m_\mathrm{Na}^2m_\mathrm{Li}}{h^2 (2m_\mathrm{Na}+m_\mathrm{Li}) \sqrt{\mu_\mathrm{NaLi}}} \nonumber \\
\times \iint \frac{rR}{\sqrt{\mu R^2 + \mu_\mathrm{NaLi} r^2}} \left[E-V(\bm{q})\right]\ dR\ d\theta,
\label{eq:rho2d}
\end{align}
where $\bm{q}=\{R, r, \theta\}$ are the Jacobi coordinates, $V(\bm{q})$ is the interaction potential, $m_x$ are the atomic masses, and $g_\mathrm{parity}=1/2$ is a factor that accounts for parity conservation.
From this we find that we should expect to encounter approximately 14 Feshbach resonances when excluding the vibrational coordinate.
This is a useful reference for the scattering calculations, where the vibrational coordinate is fixed.
When vibrations are accounted for, the number of resonances increases from 14 to 21.
Hence, the NaLi molecular vibrations contribute significantly to the density of states of the complex, but not by orders of magnitude.
Hence a large fraction of the resonances corresponds to NaLi in the vibrational ground state,
justifying freezing the vibrational coordinate in our scattering calculations.
Figure~\ref{fig:DOS}(a) shows the quasiclassical density of resonances as a function of $R_\mathrm{max}$, the upper integration limit used in evaluating Eqs.~\eqref{eq:rho3d} and \eqref{eq:rho2d}.
This shows that the resonances are supported by atom-molecule distances up to approximately 40~$a_0$,
which is considerably shorter than the range of van der Waals interactions $R_6 = (2\mu C_6)^{1/4} \approx 109~a_0$ in the system,
suggesting that in some sense the resonances are of short-range nature.
Nevertheless, at atom-molecule distances around 40~$a_0$ the electronic interaction is close to its $R^{-6}$ asymptotic form,
and the vast majority of the density of states is hence supported by the van der Waals potential.
This was previously also argued for NaK+K collisions \cite{wang:21c}, and may more generally be true of atom+molecule collisions \cite{frye:23}.
Finally, we will see that the coupling mechanism for these resonances involves the anisotropy of the atom+molecule interaction,
and hence this cannot be understood in terms of the long-range interaction which is isotropic for a molecule in its rotational ground state.

Because the density of states increases with $2J+1$, and because for $J=2$ resonances can occur with both $|\Delta M_S|=1$ and $2$,
the typical density of resonances for $J=2$ is a factor five larger than for $J=1$.
Hence we can conclude that most of the resonances observed, approximately 5 out of 6 resonances, can be assigned $J=2$ and are due to spin-spin coupling.

We also compute the density of states quantum mechanically using the same channel basis as used in the scattering calculations.
This is useful for a direct comparison to the scattering calculations.
To this end, we first compute adiabatic potential energy curves by diagonalizing the Hamiltonian excluding radial kinetic energy at every grid point.
On each adiabatic potential curve, we compute bound state wavefunctions using sinc-function discrete variable representation \cite{colbert:92}.
We record the number of bound states below the initial state threshold.
We repeat this at both ends of the magnetic field range, and the difference in the number of bound states provides an estimate for the number of resonances.
The results are shown in Fig.~\ref{fig:DOS}(b).
As can be seen, the total number of resonances increases with $N_\mathrm{max}$ truncation of the basis set,
and converges around $N_\mathrm{max}=30$.
At $N_\mathrm{max}=30$ the highest adiabatic potentials for the spin-stretched state no longer support bound states, thus the typical number of resonances converges,
although the background scattering length does not converge until much higher $N_\mathrm{max}$ due to the stronger interactions in the low-spin states.
The total number of resonances is close to the quasiclassical estimate, but not in perfect agreement with it due to the light masses and relatively weak interactions in the spin-stretched state.
The total number of resonances matches observations for a typical $B$-field scan in the lower spin-stretched state.
However, in the upper spin-stretched state we find a similar density of states, although typically a much lower number of resonances is observed.

\begin{figure*}
\begin{center}
\includegraphics[width=0.475\textwidth]{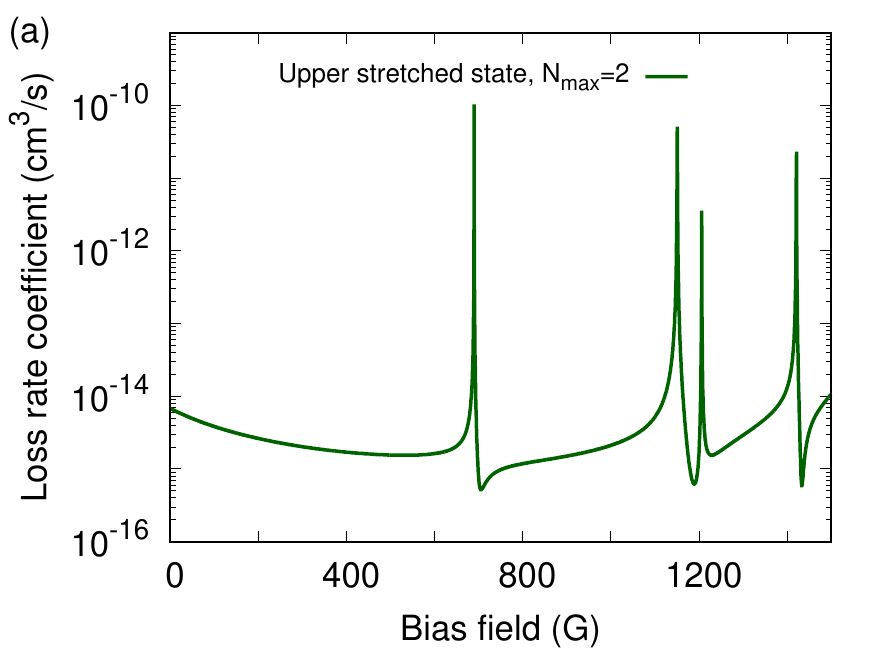}
\includegraphics[width=0.475\textwidth]{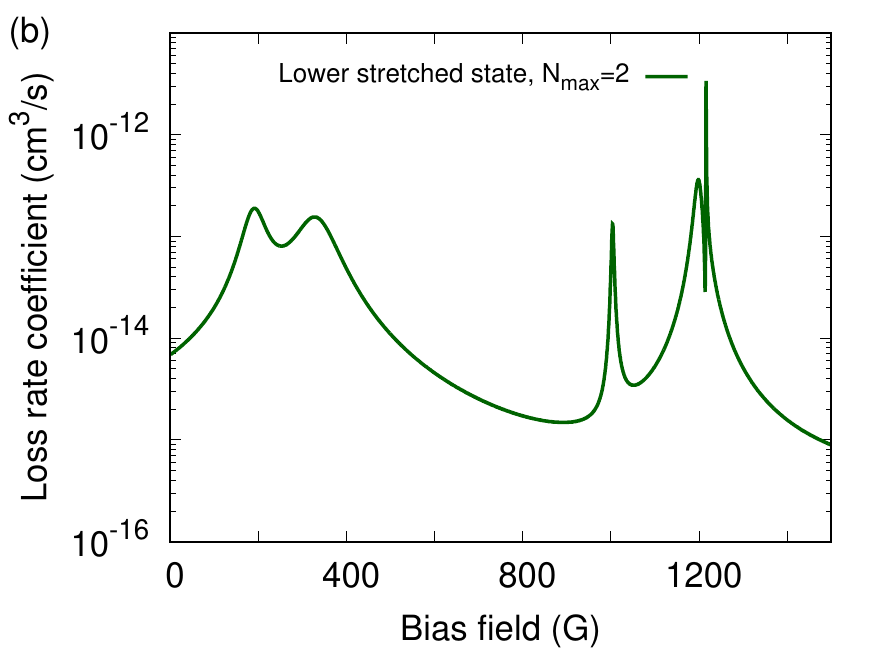}
\includegraphics[width=0.475\textwidth]{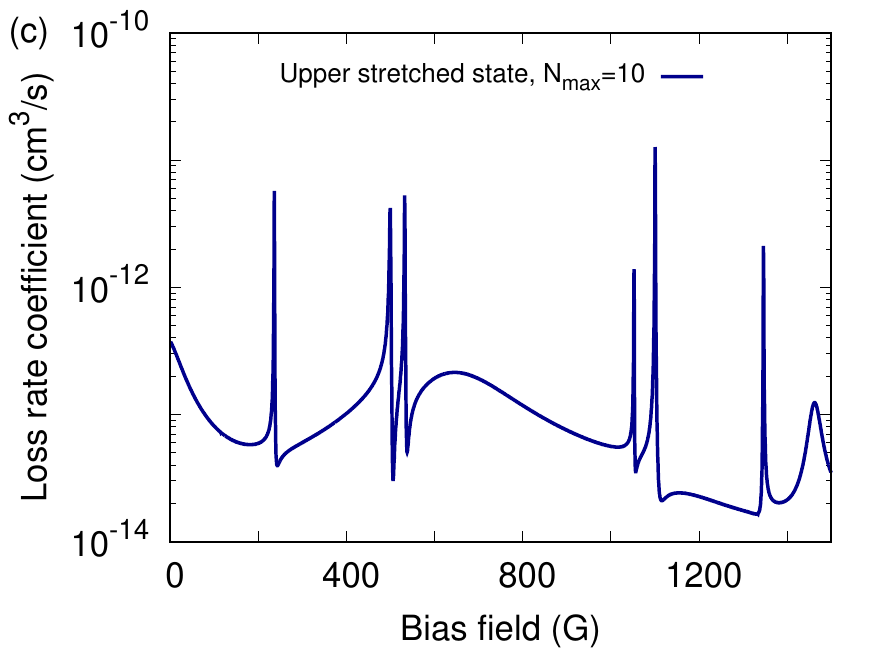}
\includegraphics[width=0.475\textwidth]{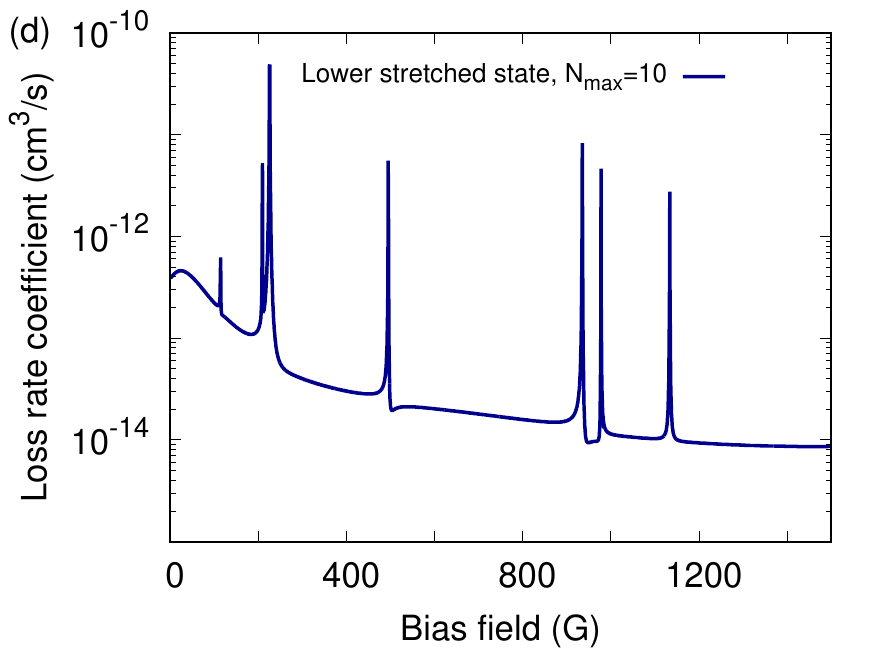}
\includegraphics[width=0.475\textwidth]{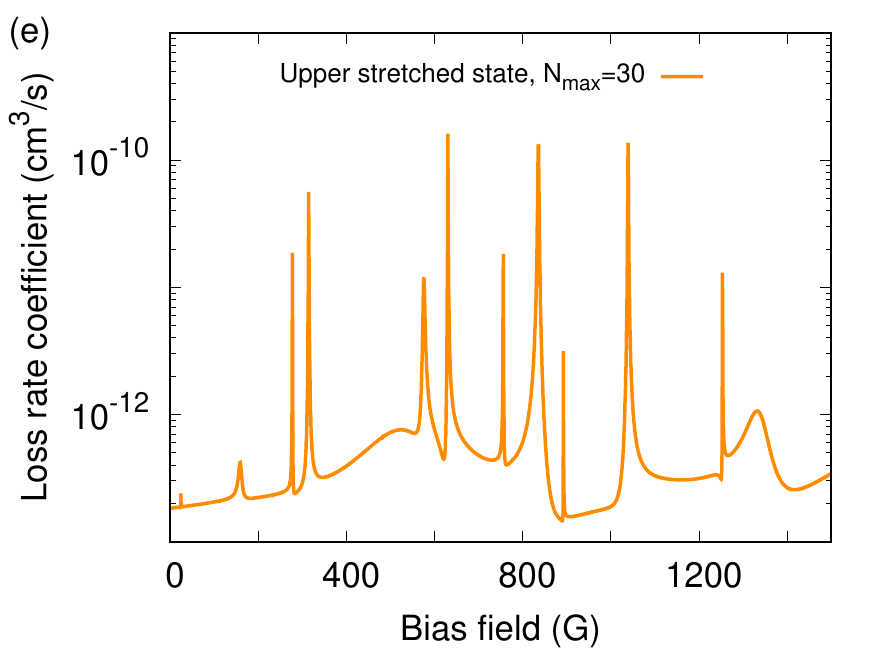}
\includegraphics[width=0.475\textwidth]{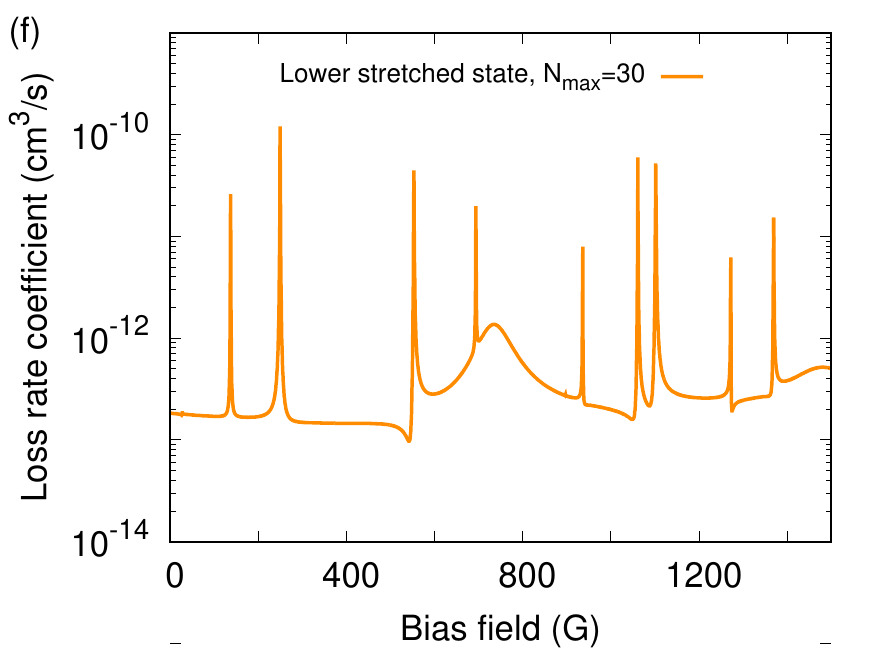}
\caption{ \label{fig:J2conv}
{\bf  Representative magnetic field scans with non-initial Zeeman states removed.}
Shown are representative magnetic field scans obtained by excluding the non-initial Zeeman states for $N=0$ from the calculation.  Different panels correspond to truncation of the basis set at $N_\mathrm{max}=2,$ 10, 30, as indicated.
The left (right) hand column shows results for the upper (lower) spin-stretched state.
Excluding the lower-lying Zeeman levels for the upper spin-stretched state increases the typical number of resonances,
whereas excluding excited Zeeman levels for the lower spin-stretched state does not reduce the typical number of resonances.
}
\end{center}
\end{figure*}

\subsection{Missing resonances in the upper spin-stretched state \label{sec:lfshfs}}

One might expect that the higher number of resonances in the lower spin-stretched state comes from quasi-bound states in the Zeeman-excited $N=0$ states.
However, this would be reflected in a higher density of states for the lower stretched state in the calculation above.
What happens is not that the total number of states is higher for the lower state,
but rather that we do not observe every state as a resonance in the upper spin-stretched state.
This is caused by fast decay of some of the resonances to the lower-lying Zeeman levels.
This is illustrated in Fig.~\ref{fig:J2conv}, which shows typical $B$-field scans from a calculation that excludes non-initial Zeeman states in $N=0$.
In the case of the lower stretched state, the typical number of resonances is not reduced by omitting the excited Zeeman levels, \emph{i.e.}, these do not cause the higher number of resonances.
In the case of the upper stretched state, the typical number of resonances is increased by \emph{omitting} the lower-lying Zeeman states.
Some of these resonances were previously not visible due to fast loss to the lower-lying Zeeman states.

Loss to lower-lying Zeeman states WITH N=0 can occur only via spin-spin coupling.  Spin-rotation coupling cannot simultaneously fulfil the parity selection rule and the selection rule $\Delta J =1$.  The argument is as follows.  In the rotational ground state $N=0$, the orbital and total mechanical angular momentum are equal, $J=L$.
For $J=1$ this means that all levels in the rotational ground state have $L=1$, and hence odd parity.
Since the parity of $N+L$ is conserved, these inelastic exit channels are inaccessible.  Note that spin-rotation coupling can lead to resonances and chemical loss since in these cases, N is changing.

\begin{figure*}
\begin{center}
\includegraphics[width=0.475\textwidth]{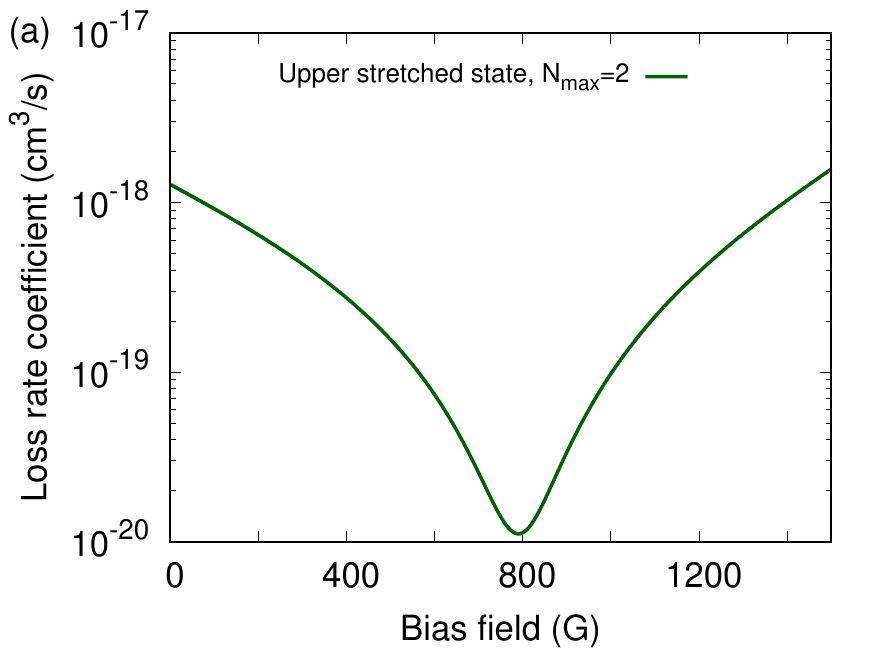}
\includegraphics[width=0.475\textwidth]{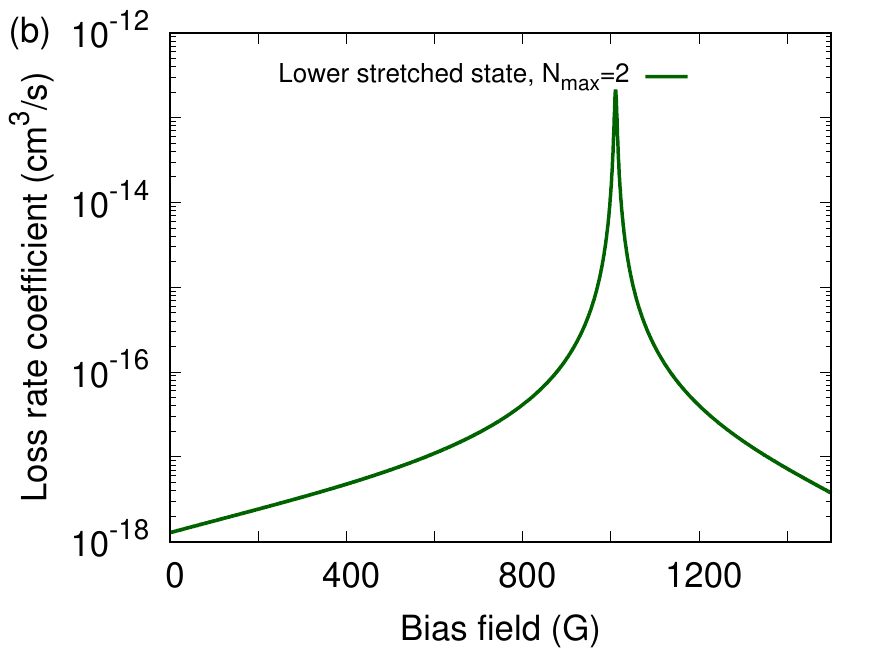}
\includegraphics[width=0.475\textwidth]{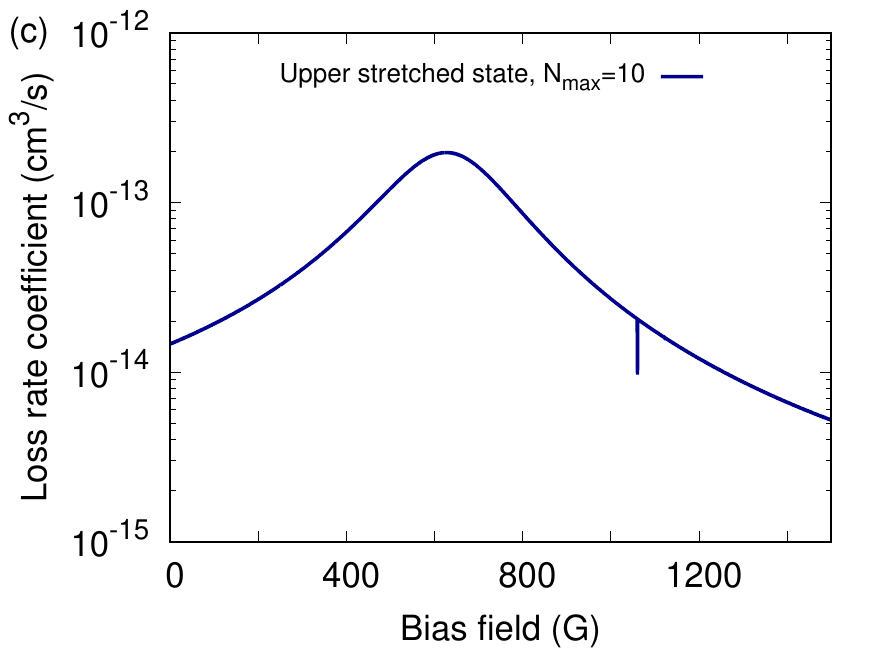}
\includegraphics[width=0.475\textwidth]{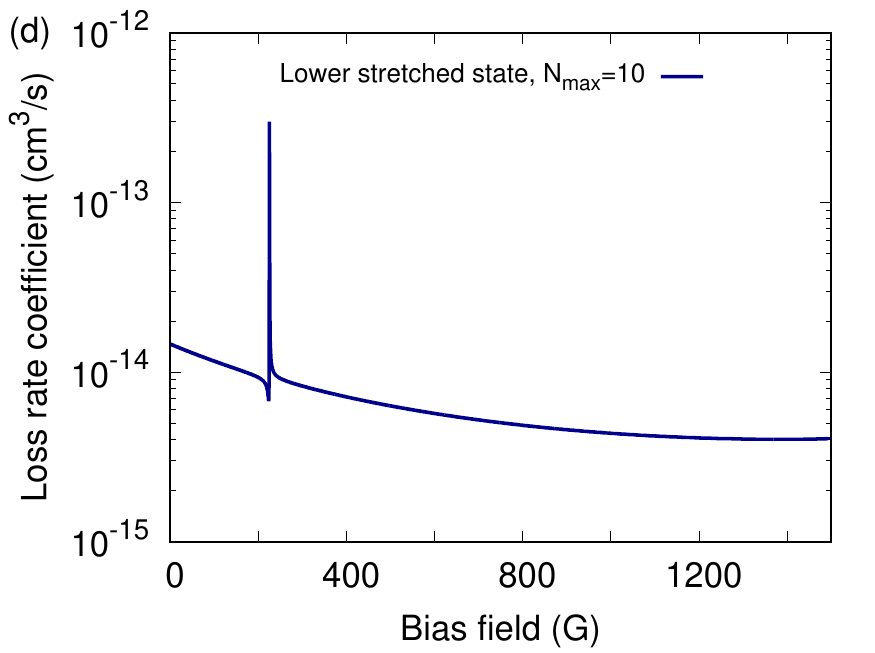}
\includegraphics[width=0.475\textwidth]{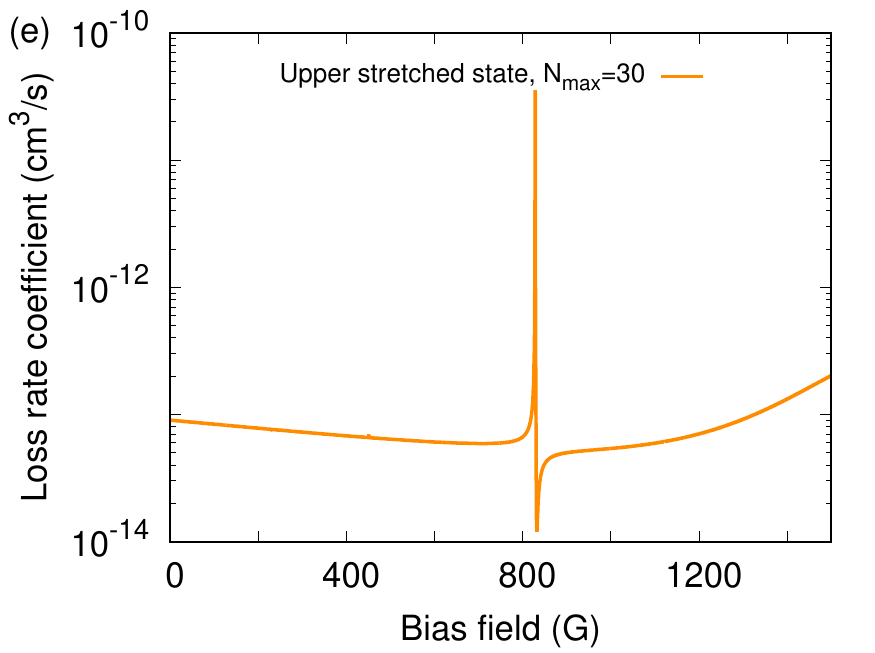}
\includegraphics[width=0.475\textwidth]{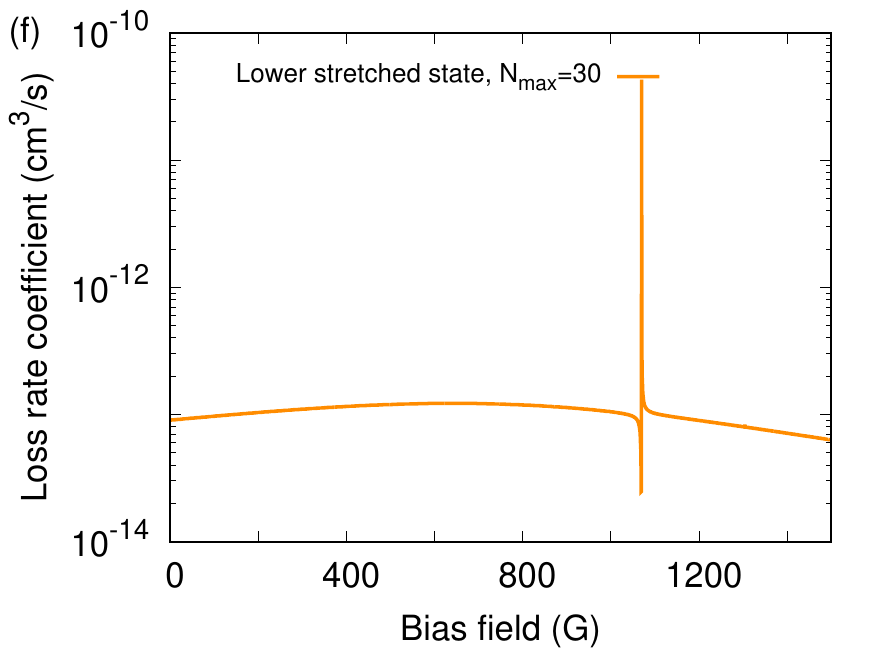}
\caption{ \label{fig:J1conv}
{\bf Representative magnetic field scans with the channel basis truncated with $J_\mathrm{max}=1$.}
Different panels correspond to truncation of the basis set at $N_\mathrm{max}=2$, 10, 30, as indicated.
The left (right) hand column shows results for the  upper (lower) spin-stretched state.
Truncation of the basis set with $J_\mathrm{max}=1$ has reduced the number of resonances,
and the qualitative difference in the number of visible resonances between the two spin states has disappeared.
}
\end{center}
\end{figure*}

Figure~\ref{fig:J1conv} shows once more typical $B$-field scans but now including $J=0,1$ states only.
This has artificially removed the contribution of the spin-spin interaction, which couples to $J=2$ states.  
Therefore, no decay to lower Zeeman states WITH N=0 is possible.  This  removes the effect discussed previously where fast inelastic scattering to lower Zeeman states renders fewer resonances observable in the upper spin-stretched state. The qualitative differences between the upper and lower spin-stretched states occur only for $J=2$ resonances, are attributed to spin-spin coupling and would therefore not occur  for collisions between atoms and spin-doublet molecules.

\subsection{Hyperfine interactions \label{sec:hf}}

\begin{figure}
\begin{center}
\includegraphics[width=0.475\textwidth]{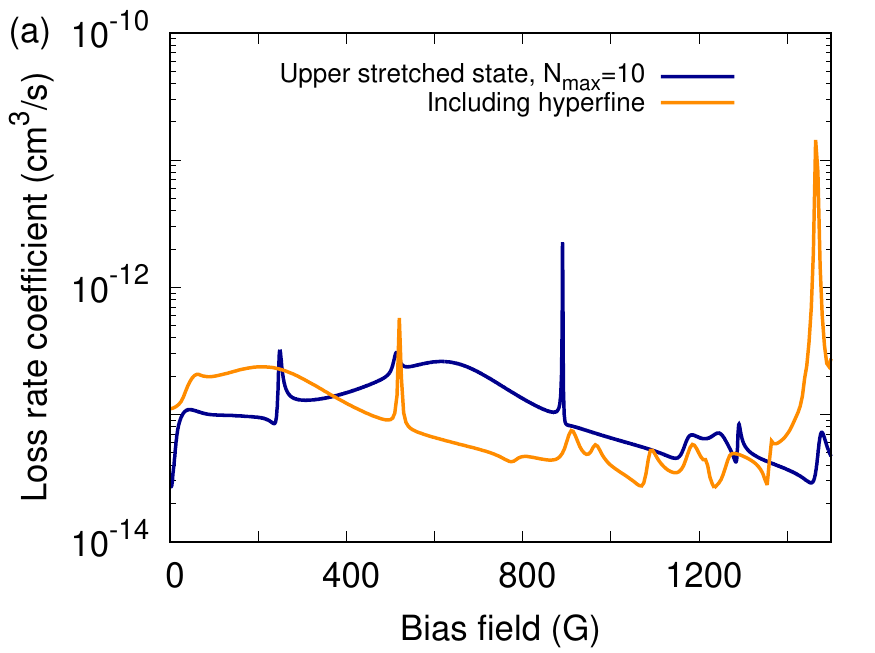}
\includegraphics[width=0.475\textwidth]{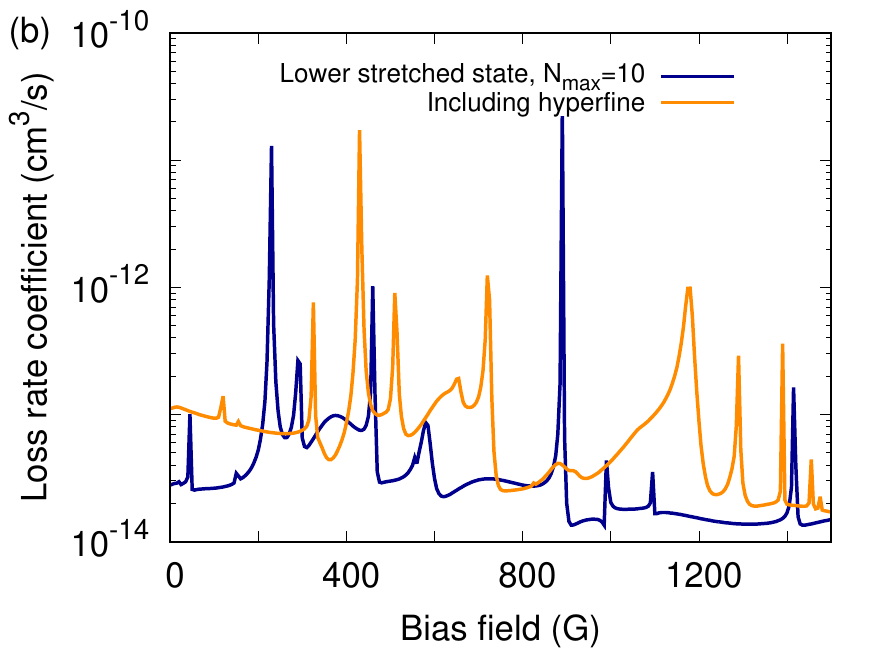}
\caption{ \label{fig:N10hf}
{\bf Including hyperfine interactions in the calcuations.}
Representative magnetic field scans are obtained with and without hyperfine interactions.
Panel {\bf (a)} shows results for the upper spin-stretched state and panel {\bf (b)} the lower spin-stretched state.
Including hyperfine interactions increases the number of resonances somewhat, but not significantly, and it does not lead to clearly identifiable multiplet splittings.
}
\end{center}
\end{figure}

Next, we investigate the effect hyperfine interactions by including nuclear spin in the coupled-channels scattering calculations.
To this end, the basis set of Eq.~\eqref{eq:basis} is extended to functions of the form
\begin{align}
|(N L) J (s~s_3) S;& \mathcal{J} \mathcal{M} \rangle|i_1 m_{i_1}\rangle|i_2 m_{i_2}\rangle|i_3 m_{i_3}\rangle.
\label{eq:hfbasis}
\end{align}
The Na nuclear spin is $i_1 = i_3 = 3/2$ and the Li nuclear spin is $i_2 = 1$.
Only functions with $\mathcal{M}+m_{i_1}+m_{i_2}+m_{i_3}$ equal to $+11/2$ are included for the upper spin-stretched state, and $-11/2$ for the lower spin-stretched state, respectively.
The hyperfine couplings included take the form $a\ \hat{i} \cdot \hat{s}$ for each atom.
We write this dot product as ${\hat{i}_z \hat{s}_z - \hat{i}_{+1} \hat{s}_{-1} - \hat{i}_{-1} \hat{s}_{+1}}$.
Matrix elements of the spherical components of the electron spin operators are given in the Appendix.
The action of the nuclear spin operators is
\begin{align}
\hat{i}_z |i m\rangle &= m|i m\rangle, \\
\hat{i}_{+1} |i m\rangle &= -\sqrt{\frac{1}{2}(i-m)(i+m+1)}\ |i, m+1\rangle, \nonumber \\
\hat{i}_{-1} |i m\rangle &=  \sqrt{\frac{1}{2}(i+m)(i-m+1)}\ |i, m-1\rangle.
\end{align}
Inclusion of hyperfine substantially increases the dimension of the basis set and we performed calculations for modest $N_\mathrm{max}=10$, shown in Fig.~\ref{fig:N10hf}.
As can be seen, the inclusion of hyperfine structure modifies the spectrum of resonances, but it does not substantially increase the number of resonances nor does it lead to clearly identifiable multiplet structure on existing resonances.

\begin{figure}
\begin{center}
\includegraphics[width=0.475\textwidth]{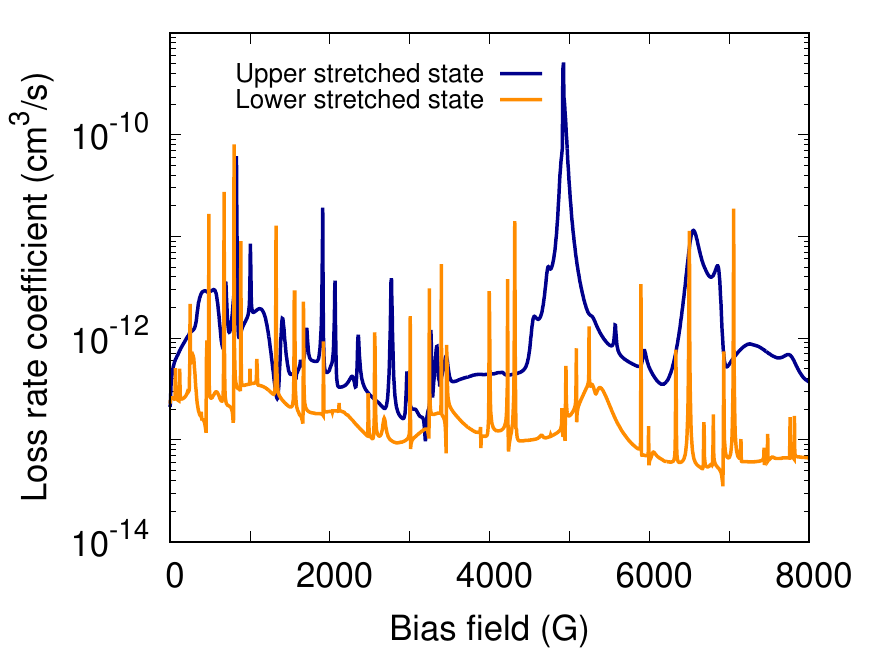}
\caption{ \label{fig:Bwide}
{\bf Collisional loss rates at higher magnetic fields.} Plotted are representative magnetic field scans up to higher magnetic fields than are probed experimentally.
In the upper spin-stretched state, a strong feature appears which is caused by a resonance between the Zeeman relaxation energy and a rotational excitation.
In the lower spin-stretched state, more resonances are typically visible,
but the strong feature near 5000~G is missing as Zeeman relaxation is not possible in this spin state.
}
\end{center}
\end{figure}

\subsection{Higher magnetic field strengths}

Finally, we consider a wider $B$-field scan of the collisional loss rate that is not accessible in the experiment.
In Fig.~\ref{fig:Bwide} we see that a strong resonance feature occurs around 5000~G for the upper spin-stretched state.
This feature occurs consistently at the same magnetic field strength and is insensitive to uncertainty of the interaction potential.
This is not a Feshbach resonance, but rather results from resonant energy transfer via spin-spin coupling where the energy released by Zeeman relaxation matches the rotational energy associated with excitation from $N=0$ to $N=2$.
This occurs only in the upper spin-stretched state, as Zeeman relaxation is not possible in the lower state.

A similar feature may be expected around 1650~G and 3300~G,
where the Zeeman energy released by a double and a single spin flip, respectively, is resonant with the $N=0$ to $N=1$ rotational transition.
This, however, is not observed as these rotational states are not coupled by spin-spin nor by spin-rotation coupling.

\subsection{Discussion}

\begin{figure}
\begin{center}
\includegraphics[width=0.475\textwidth]{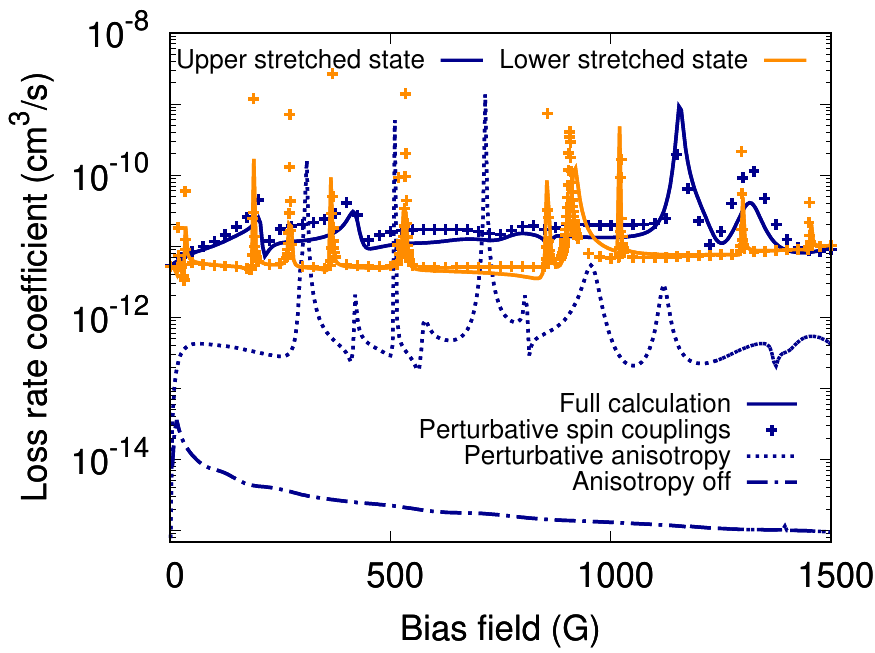}
\caption{ \label{fig:4a}
{\bf Perturbative analysis of calculated collision rates.} 
Resonance spectra are displayed for the upper and lower spin-stretched states in orange and blue solid lines, respectively,  calculated for $\lambda=-0.02$ and $N_\mathrm{max}=30$, also shown as Fig.~4(a) in the accompanying paper \cite{companion}.
Markers show four times the loss rate obtained with the spin-spin and spin-rotation couplings halved,
and agreement with the solid lines indicates these spin-dependent couplings act perturbatively.
For the upper stretched state, the dashed-dotted line indicates the loss rate obtained with the interaction anisotropy turned off, which is entirely due to the magnetic dipole-dipole interaction.
The much smaller loss rate in this case identifies a combination of the anisotropic interaction and the spin-spin and spin-rotation coupling as the dominant loss mechanism.
The dotted line is obtained by halving the strength of anisotropic interactions and multiplying the resulting loss rate by a factor four. This completely changes the shape and resonance positions. The disagreement between the dotted and the full line indicates that the anisotropic interactions do not act perturbatively.
}
\end{center}
\end{figure}

A summary of our findings is represented in Fig.~\ref{fig:4a} which can also be found in the accompanying paper \cite{companion}.
This figure compares the spectra of Feshbach resonances for the upper and lower spin-stretched states.
Markers show four times the loss rate obtained with the spin-spin and spin-rotation couplings halved.
Agreement with the solid lines indicates these spin-dependent couplings act perturbatively,
and the loss rates scale as the coupling constants squared.
Note that this does not apply to the sharp resonances, which are narrower for smaller coupling.
The dashed-dotted line indicates the loss rate obtained with the interaction anisotropy turned off.
Turning off the interaction anisotropy effectively turns off the dominant loss mechanism,
which requires the combination of spin-spin or spin-rotation coupling and the anisotropic atom-molecule interaction.
The resulting, much smaller loss rate is entirely due to the magnetic dipole-dipole interaction, and vanishes within our model if the dipole-dipole interaction is also turned off. The role of the anisotropic atom-molecule interaction, however, is not perturbative.
This can be seen from the disagreement between the solid line, which resulted from the full calculation, and the dotted line, which is obtained from a calculation where we halved the strength of anisotropic interactions and multiplying the resulting loss rate by a factor four. We also see that changing the strength of the anisotropic short-range interaction affects the resonance positions.
This means that the pattern of resonances cannot be explained in terms of the long-range \emph{isotropic} van der Waals interaction alone, as is argued in Ref.~\cite{frye:23},
despite the fact that vast majority of resonances is supported by the long-range $R^{-6}$ interaction, as argued in Ref.~\cite{frye:23} and confirmed in our density of states calculations.

In Figure~\ref{fig:4a} we also witness the qualitative difference in the number of resonances observed in the two maximally stretched spin states.
By performing calculations for various $\lambda$ scalings of the potential, we \emph{typically} observe around 5 and 10 resonances for the upper and lower spin-stretched state, respectively.
This is in qualitative agreement with experimental observation of 8 and 17 resonances in the upper and lower stretched state, respectively.
The difference between theory and experiment is partially explained by the neglect of the vibrational degree of freedom and hyperfine, as discussed above.
This difference between the two spin states in observed resonance density is not attributed to a difference in density of states,
but rather to the decay of resonances to Zeeman-relaxed channels from the upper spin-stretched state, as discussed in Sec.~\ref{sec:lfshfs}.

Figure~\ref{fig:4a} also suggests substantial differences in the background loss rate between the upper and lower spin-stretched states,
in contrast to the experimental observations where very similar loss rates were found.
Indeed, this demonstrates that for a particular realization of the calculation, for specific $\lambda$ scalings of the potential,
such differences can occur, and there is no guarantee that the two state exhibit the same background loss rate despite the losses being determined by the same mechanisms.
However, this is not necessarily the \emph{expected} behavior.
In the $\lambda$-scans of the ratio of elastic-to-inelastic collisions, shown in Fig.~\ref{fig:lambda0B},
we see that the two stretched spin-states exhibit qualitatively the same behavior,
and that for most values of $\lambda$ the loss rates are similar between the two states,
but for specific values of $\lambda$ there can be large differences.
Such large differences can occur where a resonance occurs in one of the two spin states,
but they can also reflect differences in the background scattering rates.
In Figs.~\ref{fig:Nconv}, \ref{fig:J2conv}, \ref{fig:J1conv} we see several examples where the background loss rate between the two spin states can be either similar,
different by a small factor, or different about an order of magnitude.
Where the background scattering rates are different  -- for a specific $\lambda$ scaling and basis-set truncation, $N_\mathrm{max}$ -- these differences are not systematic;
it can either be the upper or lower spin-stretched state that experiences the higher background scattering rate.
Again, as shown most clearly in Fig.~\ref{fig:lambda0B}, the \emph{expected} background behavior is similar for the two spin states.

\section{Conclusions}

We have performed coupled-channels scattering calculations of Feshbach resonances in spin-polarized NaLi $(a^3\Sigma^+)$ + Na collisions based on \emph{ab initio} interaction potentials calculated in this work.
Quantitatively predicting the background scattering length or the positions of resonances is beyond the reach of state-of-the-art theory.
However, the calculations do explain experimental observations qualitatively.
The background loss is fast in non-stretched spin states,
whereas in stretched states the ratio of elastic-to-inelastic collisions can be around 100,
in agreement with previous observations of sympathetic cooling.
When comparing the upper and lower stretched states we find the \emph{expected} background loss rate to be similar, also in agreement with experimental observations.

The calculations furthermore capture a series of Feshbach resonances.
We show that these resonance states are supported by relatively short atom-molecule separations up to $40~a_0$,
We show that the dominant coupling mechanism is a combination of the anisotropic atom-molecule interaction and the spin-spin coupling and to a lesser extent spin-rotation coupling.
The resonance states are supported by atom-molecule separations up to $40~a_0$,
where the interaction can be described by its asymptotic $R^{-6}$ form, as was previously argued for alkali-metal atom-molecule collisions \cite{wang:21c,frye:23}.
However, due to the critical role of the \emph{anisotropic} atom-molecule interaction in the coupling mechanism,
the resonance positions depend sensitively on the anisotropic short-range interactions,
and the pattern of resonances cannot be described in terms of the long-range $R^{-6}$ interaction alone.

In the lower spin-stretched state we observe approximately 10 resonances up to 1500~G.
In the upper spin-stretched state only around 5 resonances are visible due to fast decay to lower-lying Zeeman states in $N=0$.
This qualitative difference between the upper and lower spin-stretched states has also been observed experimentally,
where the two states support 8 and 17 resonances, respectively.
Molecular vibrations and hyperfine interactions, which were excluded in most of the scattering calculations that we performed, are expected to further increase the number of observable resonances.
Calculations of the density of states suggests molecular vibrations increase the number of resonances by 50~\%,
and a scattering calculation including hyperfine interaction in a small basis suggests that this too can increase the number of observed resonances somewhat. Hence, the calculations are in semi-quantitative agreement with the experimental observations.

Our combined experimental and theoretical study has shown that Feshbach resonances and collisional complexes can be well understood on the basis of state-of-the-art first principles calculations.
Due to the light elements, highly accurate electronic structure calculations can be performed on this triatomic collision complex,
with an uncertainty smaller than 3\% of the pairwise nonadditive three-body part of the potential.
To make fully quantitative predictions, the accuracy of the electronic structure calculations should be improved by more than one order of magnitude, and the coupled-channels calculations should be converged, including vibration and hyperfine structure. While it is not feasible with present-day methods and computational power, predicting the exact positions of measured Feshbach resonances in  NaLi+Na collisions constitutes a perfect testbed and playground for near-future developments of both electronic structure and scattering theory.
In the mean time, this work emphasizes we can still draw statistical conclusions regarding the density and typical width of resonances, and the typical background loss rate, that are in nearly quantitative agreement with experimental observations. Our work showcases how scattering calculations can be used as a ``\emph{numerical experiment}'' in which the interaction can be scaled at will, specific couplings can be turned off, or exit channels removed,
as a versatile tool to identify the dominant coupling mechanisms, the nature of the resonance states, and their relevant decay pathways.

\begin{acknowledgements}
M. G. and M. T. were supported by the National Science Centre Poland (Sonata Bis Grant no. 2020/38/E/ST2/00564) and the PL-Grid Infrastructure (computational grant no. PLG/2020/014342). The MIT work was supported by the NSF through the Center for Ultracold Atoms and Grant No. 1506369 and from the Air Force Office of Scientific Research (MURI, Grant No. FA9550-21-1-0069). J. P. and H. S. acknowledge additional support from the Samsung Scholarship.
\end{acknowledgements}

\newpage
\appendix
\section{Matrix elements}

\begin{widetext}
In this Appendix we give all required matrix elements in the coupled basis functions of Eq.~\eqref{eq:basis}.
First we consider the matrix elements of the electronic spin operators required for the Zeeman interaction $\hat{H}_\mathrm{Zeeman} = \mu_B g_e B (\hat{s}_{1,z}+\hat{s}_{2,z}+\hat{s}_{3,z})$.
These are
\begin{align}
\langle (N L) J (s~s_3) S;& \mathcal{J} \mathcal{M} | \hat{s}_{1,q} | (N' L') J' (s^\prime~s_3) S'; \mathcal{J}' \mathcal{M} \rangle = \nonumber \\
&\delta_{N,N'} \delta_{L,L'} \delta_{J,J'} (-1)^{2\mathcal{J}-\mathcal{M}+J+2S'+s+s^\prime+s_1+s_2+s_3+1} \left[\mathcal{J},\mathcal{J}',S,S',s,s^\prime\right]^{1/2} \nonumber \\
&\times \threejm{\mathcal{J}}{\mathcal{M}}{1}{q}{\mathcal{J}'}{\mathcal{M}'} \sixj{\mathcal{J}}{1}{\mathcal{J}}{S'}{J}{S} \sixj{S}{1}{S'}{s^\prime}{s_3}{s} \sixj{s}{1}{s^\prime}{s_1}{s_2}{s_1} \sqrt{s_1(s_1+1)(2s_1+1)},
\end{align}
and similarly
\begin{align}
\langle (N L) J (s~s_3) S;& \mathcal{J} \mathcal{M} | \hat{s}_{2,q} | (N' L') J' (s^\prime~s_3) S'; \mathcal{J}' \mathcal{M} \rangle = \nonumber \\
&\delta_{N,N'} \delta_{L,L'} \delta_{J,J'} (-1)^{2\mathcal{J}-\mathcal{M}+J+2S'+2s^\prime+s_1+s_2+s_3+1} \left[\mathcal{J},\mathcal{J}',S,S',s,s^\prime\right]^{1/2} \nonumber \\
&\times \threejm{\mathcal{J}}{\mathcal{M}}{1}{q}{\mathcal{J}'}{\mathcal{M}'} \sixj{\mathcal{J}}{1}{\mathcal{J}}{S'}{J}{S} \sixj{S}{1}{S'}{s^\prime}{s_3}{s} \sixj{s}{1}{s^\prime}{s_2}{s_1}{s_2} \sqrt{s_2(s_2+1)(2s_2+1)},
\end{align}
and
\begin{align}
\langle (N L) J (s~s_3) S;& \mathcal{J} \mathcal{M} | \hat{s}_{3,q} | (N' L') J' (s^\prime~s_3) S'; \mathcal{J}' \mathcal{M} \rangle = \nonumber \\
&\delta_{N,N'} \delta_{L,L'} \delta_{s,s'} \delta_{J,J'} (-1)^{2\mathcal{J}-\mathcal{M}+J+2S'+s+s_3} \left[\mathcal{J},\mathcal{J}',S,S'\right]^{1/2} \nonumber \\
&\times \threejm{\mathcal{J}}{\mathcal{M}}{1}{q}{\mathcal{J}'}{\mathcal{M}'} \sixj{\mathcal{J}}{1}{\mathcal{J}}{S'}{J}{S} \sixj{S}{1}{S'}{s^\prime}{s_3}{s} \sqrt{s_3(s_3+1)(2s_3+1)},
\end{align}
where the $q=-1,0,1$ spherical components of the spin operators are given by $\hat{s}_{0} = \hat{s}_z$ and $\hat{s}_{\pm 1} = \mp ( \hat{s}_x \pm i \hat{s}_y )/\sqrt{2}$.

The remaining operators we consider are all scalar, and diagonal in $\mathcal{J}$ and $\mathcal{M}$.
For the spin-rotation coupling $\gamma_s \hat{N} \cdot \hat{s}$ with $\gamma_s=0.005~$cm$^{-1}$ we need the matrix elements
\begin{align}
\langle (N L) J (s~s_3) S;& \mathcal{J} \mathcal{M} | \hat{N} \cdot \hat{s} | (N' L') J' (s^\prime~s_3) S'; \mathcal{J} \mathcal{M} \rangle = \nonumber \\
&\delta_{N,N'} \delta_{L,L'} \delta_{S,S'} (-1)^{2J'+S+\mathcal{J}+N+L+s+s_3+S'} \left[J,J',S,S'\right]^{1/2} \nonumber \\
&\times \sixj{J}{1}{J'}{N'}{L}{N} \sixj{S}{1}{S'}{s^\prime}{s_3}{s} \sqrt{N(N+1)(2N+1)s(s+1)(2s+1)}.
\end{align}
The spin-spin coupling is given by $\lambda_s \sqrt{30}/3 \left[\left[\hat{s} \otimes\hat{s}\right]^{(2)} \otimes C^{(2)}(\hat{r}_\mathrm{NaLi}) \right]^{(0)}_0$
with $\lambda_s = -0.0189~$cm$^{-1}$.
Here, 
\begin{align}
\left[ A^{(k_1)} \otimes B^{(k_2)} \right]^{(k)}_q = \sum_{q_1,q_2} A^{(k_1)}_{q_1} B^{(k_2)}_{q_2} \langle k_1 q_1 k_2 q_2 | k q \rangle
\end{align}
is the rank-$k$ spherical tensor product and $C^{(2)}(\hat{\bm{r}}_\mathrm{NaLi})$ is a tensor of Racah-normalized spherical harmonics depending on the Euler angles of the molecular axis, $\bm{r}_\mathrm{NaLi}$.
For the matrix elements we have
\begin{align}
\langle (N L) J (s~s_3) S;& \mathcal{J} \mathcal{M} | \left[\left[\hat{s} \otimes\hat{s}\right]^{(2)} \otimes C^{(2)}(\hat{r}_\mathrm{NaLi}) \right]^{(0)}_0  | (N' L') J' (s^\prime~s_3) S'; \mathcal{J} \mathcal{M} \rangle = \nonumber \\
&\delta_{L,L'} \delta_{s,s^\prime} (-1)^{2J'+S+\mathcal{J}+N+L+s+s_3+S'} \left[J,J',S,S'\right]^{1/2} \nonumber \\
&\times \sixj{J}{2}{J'}{N'}{L}{N} \sixj{S}{2}{S'}{s^\prime}{s_3}{s} \threejm{N}{0}{2}{0}{N'}{0} \frac{\sqrt{s(2s-1)}}{\sqrt{5} \threejm{s}{-s}{2}{2}{s}{s-2}}.
\end{align}

In order to evaluate the magnetic dipole-dipole interaction
$\hat{V}_\mathrm{magn. dip} = -\sqrt{30} (\mu_B g_e \alpha)^2 R^{-3} \left[\left[\hat{s} \otimes\hat{s}_3 \right]^{(2)} \otimes C^{(2)}(\hat{R}) \right]^{(0)}_0$,
we use the following
\begin{align}
\langle (N L) J (s~s_3) S;& \mathcal{J} \mathcal{M} | \left[\left[\hat{s} \otimes\hat{s}_3 \right]^{(2)} \otimes C^{(2)}(\hat{R}) \right]^{(0)}_0  | (N' L') J' (s^\prime~s_3) S'; \mathcal{J} \mathcal{M} \rangle = \nonumber \\
&\delta_{N,N'} \delta_{s,s^{\prime}} (-1)^{J'+S+\mathcal{J}+N+L'+J+L} \left[L,L',J,J',S,S'\right]^{1/2} \nonumber \\
&\times \sixj{J}{2}{J'}{L'}{N}{L} \ninej{s}{s^{\prime}}{1}{s_3}{s_3}{1}{S}{S'}{2} \threejm{L}{0}{2}{0}{L'}{0} \sqrt{s(s+1)(2s+1)s_3(s_3+1)(2s_3+1)}.
\end{align}

Finally, for the electronic interaction we have the following expressions
\begin{align}
\langle (N L) J (s~s_3) S;& \mathcal{J} \mathcal{M} | P_\ell(\cos\theta)  | (N' L') J' (s^{\prime}~s_3) S'; \mathcal{J} \mathcal{M} \rangle = \nonumber \\
& \delta_{J,J'} \delta_{s,s^{\prime}} \delta_{S,S'} (-1)^{J+S+\mathcal{J}+N+L} \left[\mathcal{J},J,J',\ell,N,N',L,L'\right]^{1/2} \nonumber \\
&\times \ninej{N}{N'}{\ell}{L}{L'}{\ell}{J}{J'}{0} \sixj{\mathcal{J}}{0}{\mathcal{J}'}{J'}{S}{J} \threejm{N}{0}{\ell}{0}{N'}{0} \threejm{L}{0}{\ell}{0}{L'}{0},
\end{align}
\begin{align}
\langle (N L) J (s~s_3) S;& \mathcal{J} \mathcal{M} | P_\ell(\cos\theta) \hat{s}_1 \cdot \hat{s}_3  | (N' L') J' (s^{\prime}~s_3) S'; \mathcal{J} \mathcal{M} \rangle = \nonumber \\
& \delta_{J,J'} \delta_{S,S'} (-1)^{N+L+s_1+s_2+s^{\prime}} \left[\mathcal{J},J,J',\ell,N,N',L,L',S,S',s,s^{\prime},1\right]^{1/2} \nonumber \\
&\times \ninej{N}{N'}{\ell}{L}{L'}{\ell}{J}{J'}{0} \ninej{J}{J'}{0}{S}{S'}{0}{\mathcal{J}}{\mathcal{J}'}{0} \ninej{s}{s^{\prime}}{1}{s_3}{s_3}{1}{S}{S'}{0} \sixj{s}{1}{s^{\prime}}{s_1}{s_2}{s_1} \threejm{N}{0}{\ell}{0}{N'}{0} \threejm{L}{0}{\ell}{0}{L'}{0} \nonumber \\
&\times \sqrt{s_1(s_1+1)(2s_1+1)s_3(s_3+1)(2s_3+1)}
\end{align}
and
\begin{align}
\langle (N L) J (s~s_3) S;& \mathcal{J} \mathcal{M} | P_\ell(\cos\theta) \hat{s}_2 \cdot \hat{s}_3  | (N' L') J' (s^{\prime}~s_3) S'; \mathcal{J} \mathcal{M} \rangle = \nonumber \\
& \delta_{J,J'} \delta_{S,S'} (-1)^{N+L+s_1+s_2+s} \left[\mathcal{J},J,J',\ell,N,N',L,L',S,S',s,s^{\prime},1\right]^{1/2} \nonumber \\
&\times \ninej{N}{N'}{\ell}{L}{L'}{\ell}{J}{J'}{0} \ninej{J}{J'}{0}{S}{S'}{0}{\mathcal{J}}{\mathcal{J}'}{0} \ninej{s}{s^{\prime}}{1}{s_3}{s_3}{1}{S}{S'}{0} \sixj{s}{1}{s^{\prime}}{s_2}{s_1}{s_2} \threejm{N}{0}{\ell}{0}{N'}{0} \threejm{L}{0}{\ell}{0}{L'}{0} \nonumber \\
&\times \sqrt{s_2(s_2+1)(2s_2+1)s_3(s_3+1)(2s_3+1)}.
\end{align}
The full expansion of the interaction is given in Eq.~\eqref{eq:expansion},
and the expansion coefficients are determined as explained in the main text.

\end{widetext}

\end{document}